\documentclass{aa} 

\usepackage{graphicx}
\usepackage{indentfirst} 
\usepackage{txfonts}
\usepackage[utf8]{inputenc}
\usepackage[T1]{fontenc}

\emergencystretch=\maxdimen
\usepackage{dcolumn} 
\usepackage{textgreek}
\usepackage{morefloats}
\usepackage{pdflscape}
\usepackage{afterpage}
\usepackage{graphicx}
\usepackage{changepage}
\usepackage{longtable}
\usepackage{threeparttable}
\usepackage{multicol}
\usepackage{multirow}
\usepackage{rotating}
\usepackage{hyperref}
\usepackage{float}
\usepackage{txfonts}
\usepackage{siunitx}
\usepackage[Symbolsmallscale]{upgreek}
\usepackage{fancyvrb,cprotect}

\DeclareMathOperator\erf{erf}

\usepackage{cleveref}
\crefname{section}{§}{§§}
\Crefname{section}{§}{§§}

\usepackage{color, colortbl}
\definecolor{Gray}{gray}{0.9}
\definecolor{Green}{rgb}{0,1,0}
\definecolor{Red}{rgb}{1,0,0}
\usepackage[table]{xcolor}

\usepackage{hyperref}
\usepackage{xcolor}
\hypersetup{
  colorlinks   = true, 
  urlcolor     = black, 
  linkcolor    = blue, 
  citecolor    = blue 
}
\usepackage[all]{hypcap} 

\usepackage{comment}

\begin{document}

  \title{The Three Hundred project: Radio luminosity evolution from merger-induced shock fronts in simulated galaxy clusters}
  \titlerunning{Merger shocks and radio relics in the Three Hundred galaxy cluster sample}

  \author{
    S.~E. Nuza \inst{1} \and
    M. Hoeft \inst{2} \and 
    A. Contreras-Santos \inst{3,4} \and
    A. Knebe \inst{3,4,5} \and
    G. Yepes \inst{3,4} 
    }

  \institute{
    Instituto de Astronomía y Física del Espacio (IAFE, CONICET-UBA), CC 67, Suc. 28, 1428 Buenos Aires, Argentina\\
    \email{snuza@iafe.uba.ar}
    \and
    Thüringer Landessternwarte, Sternwarte 5, 07778 Tautenburg, Germany 
    \and
    Departamento de F\'isica Te\'{o}rica, M\'{o}dulo 15, Facultad de Ciencias, Universidad Aut\'{o}noma de Madrid, 28049 Madrid, Spain
    \and
    Centro de Investigaci\'{o}n Avanzada en F\'isica Fundamental (CIAFF), Facultad de Ciencias, Universidad Aut\'{o}noma de Madrid, 28049 Madrid, Spain
    \and
    International Centre for Radio Astronomy Research, University of Western Australia, 35 Stirling Highway, Crawley, Western Australia 6009, Australia
    }

   \date{Received XXX; accepted YYY}

   \abstract{
     Galaxy cluster mergers are believed to generate large-scale shock waves that are ideal sites for cosmic ray production. In these so-called radio relic shocks, synchrotron radiation is produced mainly as a result of electron acceleration in the presence of intracluster magnetic fields.   
     }{
     We aim to compute radio emission light curves for a sample of galaxy group and cluster mergers simulated in a cosmological context in order to study the dependence of radio luminosity on cluster mass, redshift, and impact parameter. 
     }{
     We used model galaxy clusters from \textsc{The Three Hundred} project, a sample of 324 simulated high-density regions of radius $15\,h^{-1}\,$Mpc extracted from a cosmological volume, to identify cluster mergers characterised by the two main merging structures, construct their light curves, and follow their evolution throughout the complete simulated cosmic history.  
     }{
     We found that the median non-thermal radio relic luminosity light curve produced in galaxy cluster mergers can be described by a skewed Gaussian function abruptly rising after core-passage of the secondary cluster that peaks after $\sim0.1$--$0.8\,$Gyr as a function of $M_{200,1}$, the mass of the primary, displaying a mass-dependent luminosity output increase of $\lesssim10$ to about $\gtrsim10$--$50$ times relative to the radio emission measured at core-passage for galaxy groups and clusters, respectively. In general, most merger orbits are fairly radial with a median opening angle of $\sim20^{\circ}$ before the collision. We also found that, independent of the cluster mass, less radial mergers tend to last longer, although the trend is weak. Finally, in agreement with previous works, we found that the peak radio luminosity shows a significant correlation with mass, $P_{1.4}\propto M_{200,1}^{2.05}$, demonstrating that this relation holds all the way up from galaxy group scales to the most massive galaxy clusters. 
     }{
     We conclude that cluster mass is the primary driver for radio `gischt' median luminosity, although there are significant variations for a given cluster mass related to the specifics of each merging process. In general, binary mergers are responsible for many of the well-known observed radio relic structures but complex situations involving three or more substructures are also common. Our simulations suggest that the shock-driven, non-thermal radio emission observed on cluster outskirts are the result of massive galaxy cluster mergers at $z\lesssim1$, peaking at $z\sim0$--$0.5$.
     }
    


   \maketitle
%

\section{Introduction}

  %
  Galaxy clusters are the most massive gravitationally bound structures in the Universe. They grow by cosmological mass accretion and mergers with other structures \citep{2012ARA&A..50..353K}. During their lifetime, massive clusters typically undergo a small number of major mergers -- collisions of two similarly massive clusters -- until they reach their present size. Clusters are also an extremely powerful probe of cosmology since their growth is essentially driven by gravity and depends on the initial perturbation spectrum in the density and velocity fields of the cosmic matter distribution \citep{2017MNRAS.471.1370S, Abdullah24, Qiu24}. 

  %
  In order to constrain cosmological parameters from the galaxy cluster distribution, it is crucial to determine their mass. Mass estimations, such as those derived with the Sunyaev-Zel'dovich method \citep[e.g.][]{2018A&A...610L...4H}, can suffer from systematic biases, so accurate calibration is crucial. Hydrostatic masses in relaxed clusters are a cornerstone in galaxy cluster studies, although several effects such as clumpiness in the intracluster medium (ICM) -- the thin, magnetised plasma that fills the space between cluster galaxies -- bulk gas motions, turbulence, and the presence of non-thermal components, have an impact on the pressure profile, leading to so-called hydrostatic bias. Therefore, understanding the processes in the cluster outskirts is crucial for quantifying these effects. 

  %
  Major and minor mergers can generate large shock waves propagating in the ICM into the cluster periphery \citep{Lokas23}. These shock fronts have been confirmed as jumps in X-ray surface brightness and temperature profiles that abruptly change the local ICM properties, affecting its evolution (see e.g. \citealt{2013ApJ...764...82B,2015MNRAS.449.1486S,2016MNRAS.463.1534B} for the case of merger shocks in the Abell\,521, Bullet, and El Gordo clusters, respectively). For instance, dissipation of kinetic energy at merger shocks heats up the plasma, whereas their megaparsec-size curvature may lead to turbulence in the ICM at smaller scales. 

  %
  Diffuse radio emission features, so-called `radio relics' or `gischts', have been found on the outskirts of galaxy clusters \citep{2019SSRv..215...16V}. These structures are usually large-scale, elongated radio features believed to trace merger shocks \citep{1998A&A...332..395E,HB07}. For some relics, a clear connection to the discontinuity in the ICM has been made (see \citealt{Wittor21} for a compilation of relics with associated merger shocks discovered in the X-ray band). Generally, relics are found in merging galaxy clusters, as has recently been shown in the relic compilation of the LOFAR two-metre Sky Survey \citep{Jones23}, indicating that radio relics occur only after a merger, when clusters are still dynamically perturbed.  

  %
  A `standard model' for the origin of radio relics has been developed \citep{1998A&A...332..395E,HB07},  according to which shock fronts accelerate electrons to relativistic energies by diffusive shock acceleration (DSA; e.g. \citealt{Drury83}), and the observed radio emission results from synchrotron radiation of cosmic ray electrons (CRes) cooling downstream of the shock. This scenario has been implemented in cosmological, (magneto-)hydrodynamical simulations and a striking similarity between simulated and observed relics has been found \citep[][]{Hoeft08,Hoeft11,Skillman11,Nuza11,Nuza12,Gelszinnis15,Nuza17,Ha18,2019MNRAS.490.3987W,Boess23,Lee24}. 
  
  In particular, \citet{Nuza17} compared a sample of elongated relics drawn from the MUSIC-2 galaxy cluster simulations \citep{Sembolini13} with observations and found that, according to the standard model, the illumination of merger shocks reproduces properties of observed relics -- such as location, age, and morphology -- very well. However, according to the same scenario, a very high, or even nonphysical, conversion of thermal energy into CRe energy might be required to reproduce the observed radio luminosities \citep{2020A&A...634A..64B}. In this respect, a plausible explanation for the latter invokes the (re-)acceleration of pre-existing plasma \citep{Pinkze13}, but it remains unclear if this would affect all merger shocks in a similar way, keeping the radio illumination of the shock fronts uniform.

  %
  Based on simulated clusters extracted from the {\sc MareNostrum} cosmological simulation, \citet{Nuza12} introduced a radio luminosity probability distribution for clusters in order to parametrise the radio power of relics as a function of the observed frequency, cluster mass, and redshift. In \citet{2020MNRAS.493.2306B} and \citet{2022MNRAS.517.1299Z}, the radio relic luminosity was related to the cluster merger probability based on an extended Press-Schechter model. Their semi-analytical relic model assumes a uniform shock front propagating at a typical shock velocity to the cluster periphery. The radio luminosity is modulated according to the local plasma properties. Finally, an ad hoc assumption is made that the shock front starts at half the virial radius.  

  %
  In this paper, we use a large sample of hydrodynamical simulations of galaxy clusters belonging to the \textsc{The Three Hundred} project \citep{Cui18} to identify major galaxy cluster mergers and model the evolution of the associated merger shocks to study the evolution of radio luminosity with time; that is, the radio relic light curve just before the core-passage to the demise of the gischt emission. Since radio relics are believed to be caused by synchrotron losses, the final radio output mainly depends on the CRe density and the strength of the magnetic field inside the relic volume. The CRe density depends, in turn, on the strength of the shock, characterised by the sonic Mach number, $\mathcal{M}$. This generates a modulation of the radio luminosity with local properties as shocks propagate from the central region of the cluster to its outskirts.

  The \textsc{The Three Hundred} set of simulations is able to capture local gas properties reasonably well, without the need for any ad hoc assumptions besides the magnetic field strength, for a large set of realistic cosmological merger scenarios. Given the large variety of possible radio light curves, we aim to parametrise the median radio output during galaxy group and cluster mergers as a function of the cluster mass, the merger mass ratio, and the impact parameter.

  The paper is structured as follows. In Section~\ref{sec:simulations}, we briefly present the set of galaxy cluster simulations analysed. In Section~\ref{sec:analysis}, we describe the halo identification method, the cluster merger sample, the non-thermal radio emission model, and the shock detection scheme used. In Section~\ref{sec:results}, we present the main results of this work, including the median radio luminosity evolution of the relic sample as a function of cluster mass and redshift, together with the gischt radio power and duration-mass correlations. Finally, in Section~\ref{sec:summary} we summarise our results.

\section{Simulations}
\label{sec:simulations}

The so-called \textsc{The Three Hundred} project\footnote{\url{https://the300-project.org}} consists of $324$ spherical zoom-in re-simulations of galaxy cluster regions of radius $15\,h^{-1}\,$Mpc extracted from the DM-only MDPL2 MultiDark simulation \citep{Klypin16}. The latter follows the evolution of cosmological structure until the present time in a periodic box of $1\,$cGpc\footnote{The `c' refers to comoving coordinates.} on a side containing $3840^3$ DM particles of mass $1.5\times10^9\,h^{-1}\,$M$_{\odot}$ and a Plummer equivalent softening length of $6.5\,h^{-1}\,$kpc. The cosmological model adopted in MDPL2 is consistent with the {\it Planck} 2015 cosmology \citep{Planck16}. From this simulation, the 324 most massive virialised structures were selected and re-simulated, including full hydrodynamics and other relevant baryonic processes. The DM particles at the beginning of the re-simulations were split in DM and gas particles, according to the cosmological baryon fraction, with an initial mass resolution of $1.27 \times 10^9\,h^{-1}\,$M$_{\odot}$ and $2.36 \times 10^8\,h^{-1}\,$M$_{\odot}$, respectively. In this way, clusters at $z=0$ located at the centre of the re-simulated regions comprise a mass-complete sample with masses in the range of $M_{200} = 6.4 \times 10^{14}\,h^{-1}\,$M$_{\odot}$ to $2.65 \times 10^{15}\,h^{-1}\,$M$_{\odot}$, where $M_{200}$ is the mass contained within a sphere of radius, $R_{200}$.\footnote{$R_{200}$ is the radius enclosing an overdensity of 200 times the critical density of the Universe.} The mean and median $R_{200}$ values of the central clusters in our sample at $z=0$ are $1.56\,h^{-1}\,$Mpc and $1.523\,h^{-1}\,$Mpc, respectively, with a corresponding enclosed mass of $M_{200}=9.16\times10^{14}\,h^{-1}\,$M$_{\odot}$ and $8.22\times10^{14}\,h^{-1}\,$M$_{\odot}$. 

These re-simulated volumes represent regions with radius of about $7R_{200}$ around typical sample clusters at the present time. The outer layers of the re-simulated regions (i.e. beyond $15\,h^{-1}\,$Mpc from the centre) are populated with lower-resolution particles to mimic any large-scale tidal field affecting the cluster evolution without increasing the computational cost. The simulation runs are stored in $129$ snapshots ranging from $z=16.98$ to $z=0$ and have been used to model galaxy cluster environments to perform studies of galaxy formation in high-density and large-scale filamentary regions, the hydrostatic mass bias dependence of clusters, the evolution of their dynamical state, and the thermodynamical properties of filaments feeding clusters \citep[e.g.][]{Arthur19, Haggar20, DeLuca21, Contreras22,Kuchner22,Gianfagna23,Rost24}, among other projects. 

\begin{figure}
    \hspace{-0.4cm}
    \includegraphics[width=0.97\columnwidth]{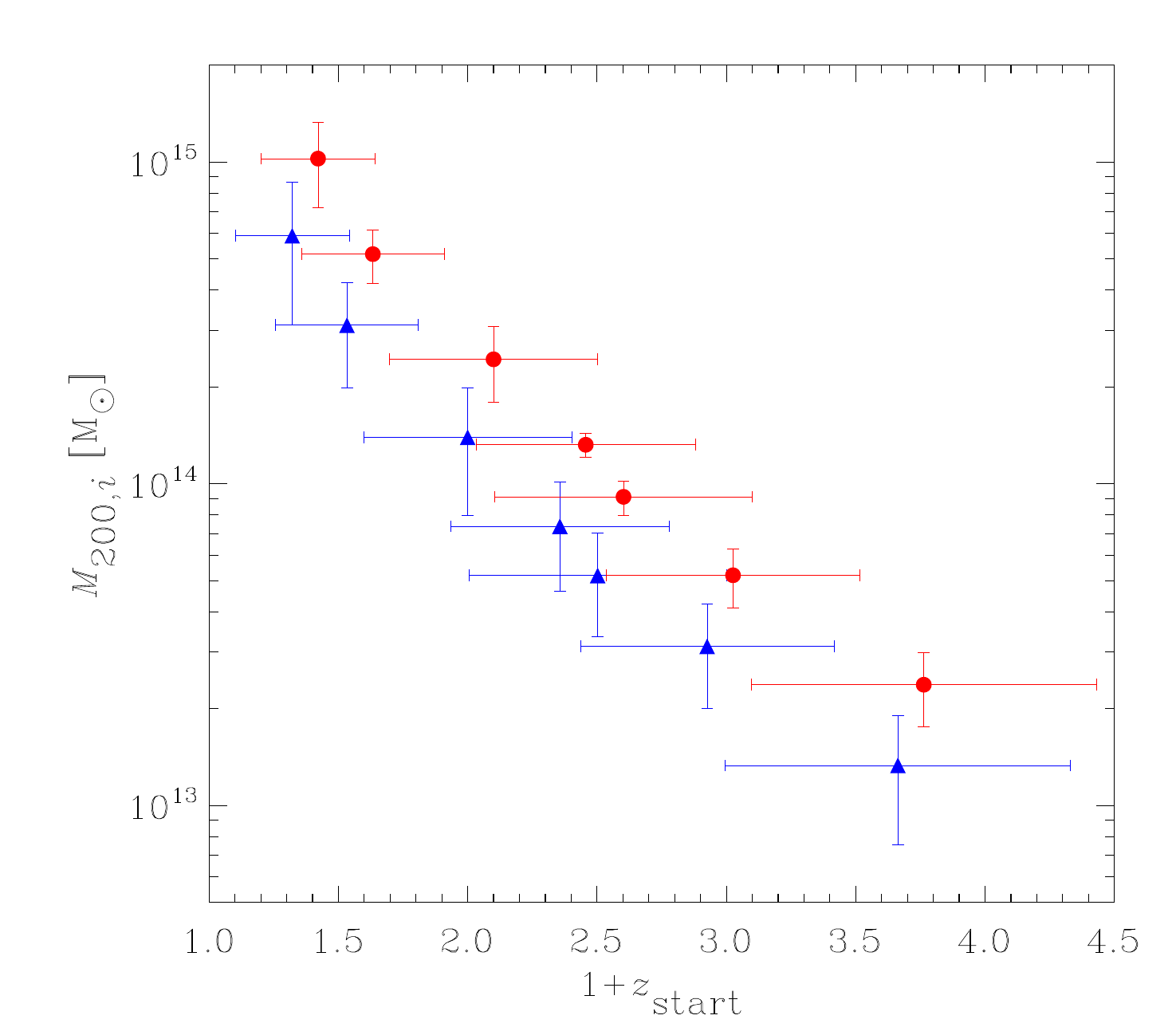}
    \caption{Mean progenitor mass of the merging groups and clusters as a function of mean redshift at $t_{\rm start}$. The solid red circles (solid blue triangles) correspond to the primary (secondary) progenitor of mass  $M_{200,1}$ ($M_{200,2}$). The error bars indicate standard deviations in each bin. For clarity, triangles have been horizontally shifted $0.1$ to the left to avoid overlap.}
    \label{fig:M200_vs_z}
  \end{figure}

The re-simulations were run with the \textsc{Gadget-X} code \citep{Murante10,Rasia15,Beck16}, which is a modified version of the non-public \textsc{Gadget3} Tree-PM gravity solver built upon the traditional \textsc{Gadget} cosmological code \citep[][]{Springel05}, and implements full gas physics within the smoothed particle hydrodynamics (SPH) approach including an improved treatment of gas mixing and gas-dynamical instabilities. The latter was achieved by adding a suitable artificial thermal diffusion to the simulations \citep[see e.g.][]{Rasia15,Beck16}. The code also includes sub-grid physical modules to treat radiative losses and other relevant astrophysical phenomena shaping galaxy evolution such as metal-dependent cooling, an homogeneous ultraviolet background, star formation, supermassive black hole growth, and feedback from supernovae and active galactic nuclei (AGNs) \citep[see e.g. Table 2 in][and references therein]{Cui18}. Throughout this paper, when reporting cluster masses and distances, we explicitly use $h=0.678$ for the dimensionless Hubble constant parameter.

\section{Analysis}
\label{sec:analysis}

\subsection{Halo identification}
\label{sec:AHF}

At each snapshot of the simulations, haloes were identified using the open-source AHF halo finder \citep[][]{Gill04,Knollmann09}, which takes into account both dark matter and baryonic components. The code searches for spherical overdensities in the density field that fall below a value of $\Delta\times\rho_{\rm crit}$, where $\Delta$ is the density contrast and $\rho_{\rm crit}$ is the critical density of the Universe at the given redshift, and determines the corresponding $R_{\Delta}$ radius in an iterative way. Typically, $R_{200}$ is used to determine halo size, whereas $\Delta\approx98$ corresponds to the density contrast of a virialised halo \citep{Bryan98}. Accordingly, halo masses for a given density contrast, $M_{\Delta}$, were computed. Additionally, for each halo, substructures within $R_{200}$ were identified in the same way and classified as subhaloes linked to the more massive host halo.  

\subsection{Cluster merger sample}
\label{sec:merger_sample}

Following \cite{Contreras22}, we built a catalogue of cluster mergers characterised by a significant mass increase of the primary merging cluster within a relatively short period, which we consider to be half of the cluster dynamical time or less, in contrast to the slower accretion of low-mass substructures during the whole galaxy cluster evolution history. As a result, we obtained a cluster sample including different merger scenarios, generally determined by two main merging systems, but also including more complex situations like multiple group or cluster mergers and the substructure accretion naturally expected in a hierarchical formation scenario.

For all merger events, we defined four characteristic times (or their corresponding redshifts) during the complete merger phase: (i) the time before the merger when the cluster is still in equilibrium ($t_{\rm before}$), (ii) the onset of the merger itself when the main progenitor starts growing according to a threshold in the fractional mass change ($t_{\rm start}$), (iii) the end of the merger when the cluster starts relaxing again ($t_{\rm end}$), and (iv) the time marking the end of the whole merger phase when the cluster has relaxed ($t_{\rm after}$). At each timestep, the dynamical state of clusters was determined using the so-called `relaxation parameter', $\chi_{\rm DS}(t)$, introduced by \cite{Haggar20}, which measures the departure from equilibrium during the lifetime of a cluster, whereby an increasing (decreasing) relaxation parameter means that the cluster is in the process of relaxing (being perturbed).

\begin{figure*}
    \centering
    \includegraphics[width=\columnwidth]{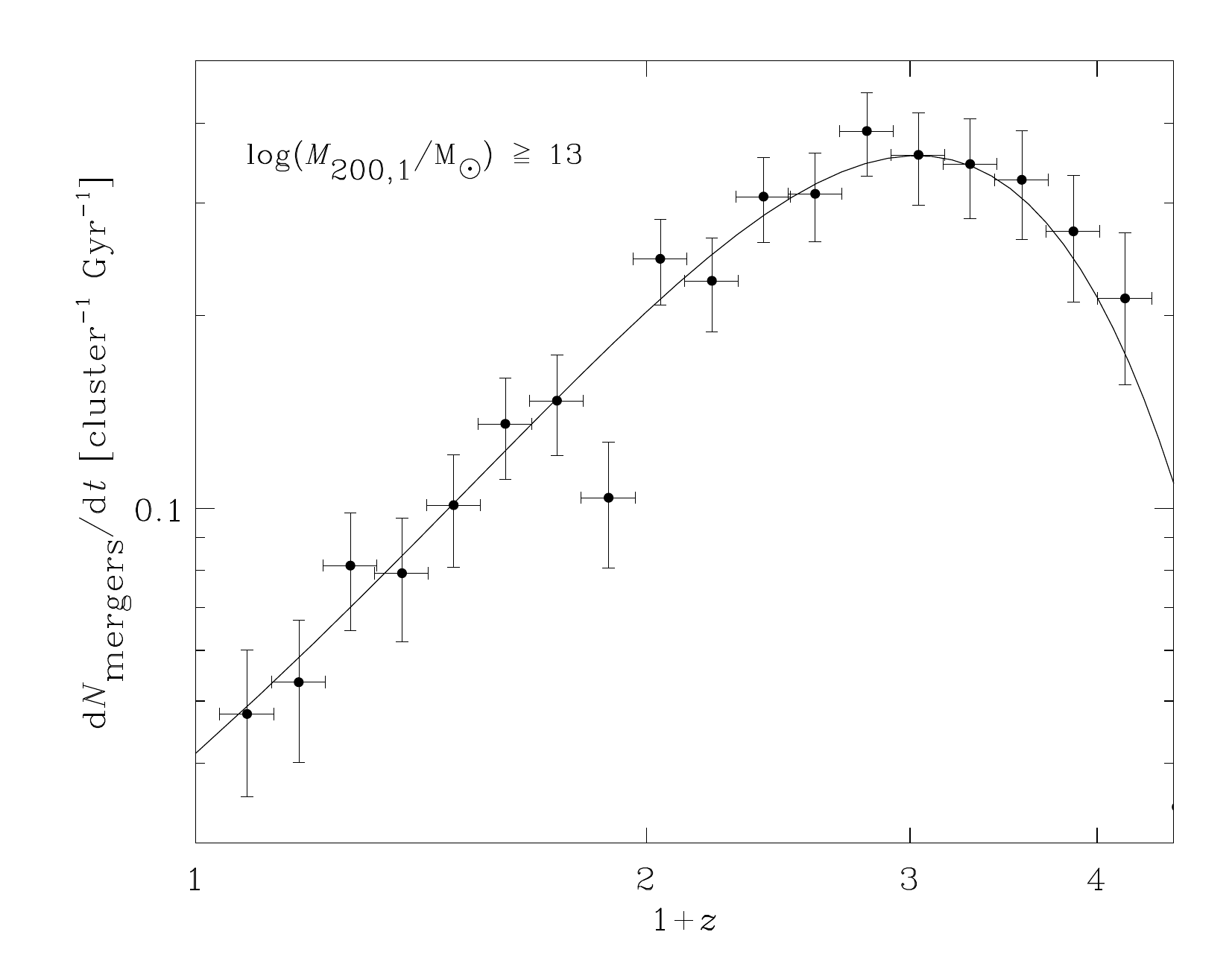}
    \includegraphics[width=\columnwidth]{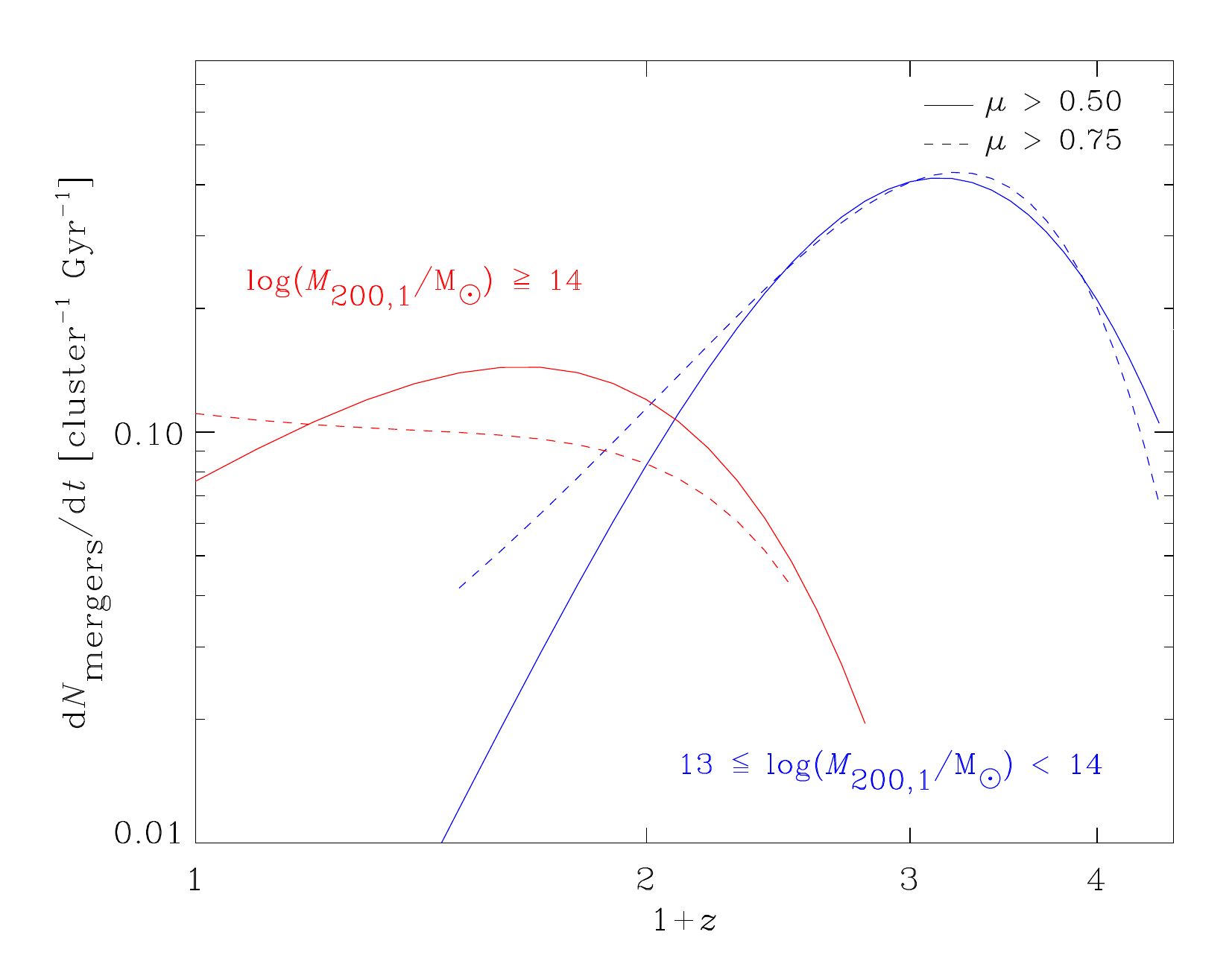}
    \caption{Galaxy cluster merger rate as a function of redshift normalised by the corresponding number of synthetic cluster regions in each mass bin (see Section~\ref{sec:merger_rate}) for major mergers with $\Delta M/M\geq 0.5$. Best-fit curves to simulated data are shown. {\it Left panel:} Data points correspond to progenitor cluster masses with $M_{\rm 200,1}\geq10^{13}\,$M$_{\odot}$ assuming Poissonian errors. {\it Right panel:} Best-fit curves for progenitor cluster masses with $10^{13}\leq M_{\rm 200,1}/{\rm M}_{\odot} < 10^{14}$ (blue) and $M_{\rm 200,1}\geq10^{14}\,$M$_{\odot}$ (red). Data points are excluded for clarity. Solid and dashed lines indicate two different cuts of the merger ratio parameter, $\mu$.}
    \label{fig:merger_rate}
\end{figure*}

The merger parameter, $\mu$, is usually defined as the mass ratio of the two most massive colliding systems; that is, $\mu \equiv M_{200,2}/M_{200,1}$, where $M_{200,1}$ and $M_{200,2}$ correspond to the largest and smallest galaxy cluster masses, which we take as $M_{200,i}\equiv M_{{200},i}(t_{\rm start})$, as we define all progenitor masses at $t_{\rm start}$. From this quantity, it is customary to classify collisions into minor ($0.1<\mu\leq0.33$) and major mergers ($\mu>0.33$).
For all clusters in \textsc{The Three Hundred} sample, we searched for an increase of at least $50\%$ in the main progenitor cluster mass at the beginning of the merger. Therefore, we demanded that the fractional mass change at $t_{\rm start}$ be $\Delta M/M\geq 0.5$. This condition ensures that most of the mergers in the catalogue will correspond to major mergers, as is shown in \cite{Contreras22}. This is reasonable, since the main branch halo is increasing its mass by at least $50\%$, and thus the secondary one is expected to be significantly massive in relation to the primary, while the remaining mass comes from minor mergers and accretion.

To make sure that all mergers were well resolved, only haloes with a minimum of $N_{\rm cut}=5000$ particles were included in the merger catalogue. In total, we ended up with a sample containing 555 mergers with $M_{200,1}\geq10^{13}\,$M$_{\odot}$. In particular, the main progenitor masses in the sample at $t_{\rm start}$ span almost two orders of magnitude across different cosmic epochs ranging from galaxy groups to large clusters of about $10^{15}\,$M$_{\odot}$. This can be seen in Fig.~\ref{fig:M200_vs_z}, where the redshift evolution of primary and secondary progenitor masses at $t_{\rm start}$ are shown. In each bin, mass and redshift distributions are characterised by their mean values and standard deviations. As was expected, more massive mergers in our catalogue take place at lower redshifts. Finally, we also note that, owing to the growth of the structure, cluster regions will end up with even larger masses at $z=0$ (see Section~\ref{sec:simulations}).

\begin{figure*}
    \begin{flushleft}
    \hspace{1.86cm}$\rho_{\rm gas}\,$\hspace{3.1cm}$kT$\hspace{3cm}$\mathcal{M}$\hspace{3.2cm}{\it KF}\hspace{2.9cm}$P_{1.4}$
    \vspace{-0.2cm}
    \end{flushleft}
    \centering
    \includegraphics[width=1.95\columnwidth]{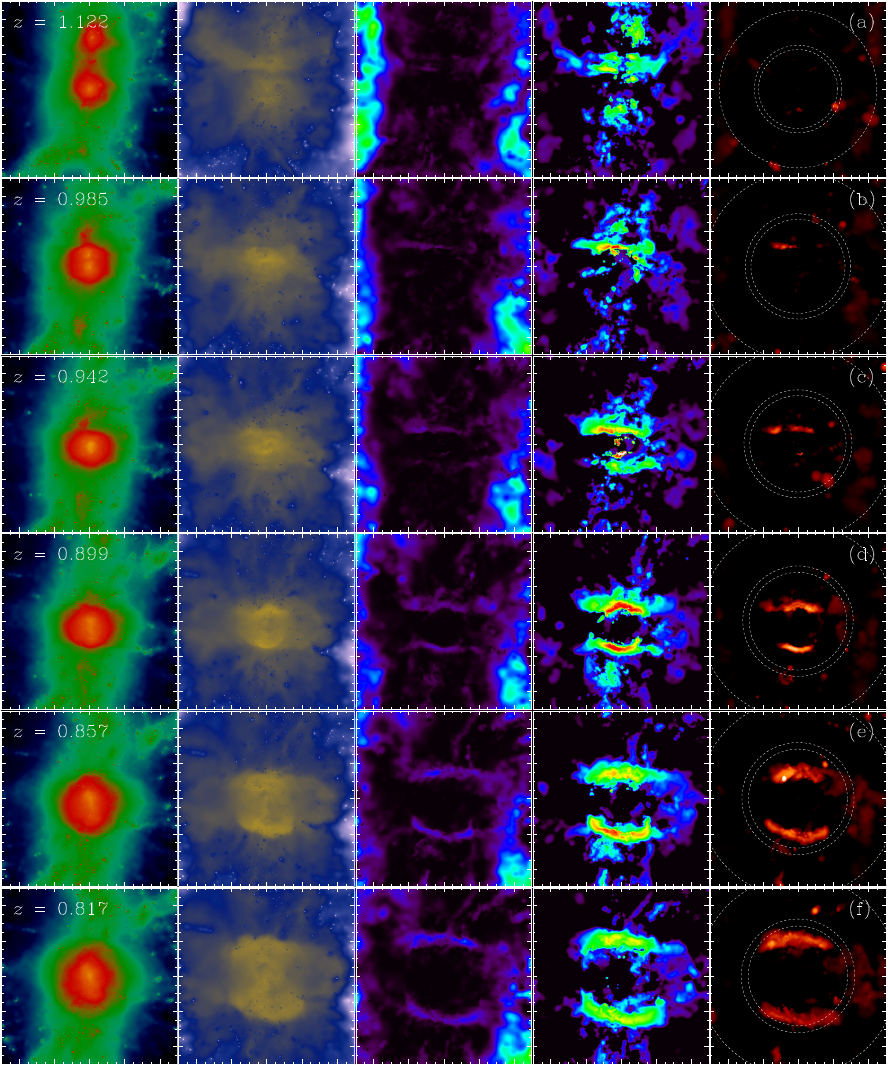}\vspace{0cm}
    \includegraphics[width=0.5\columnwidth]{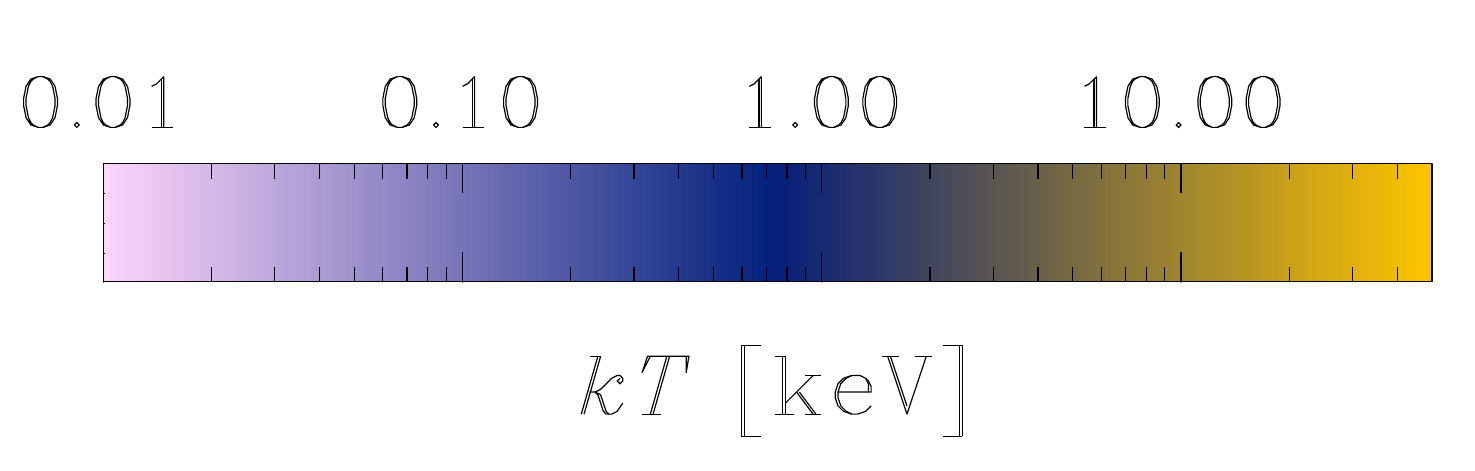}\includegraphics[width=0.5\columnwidth]{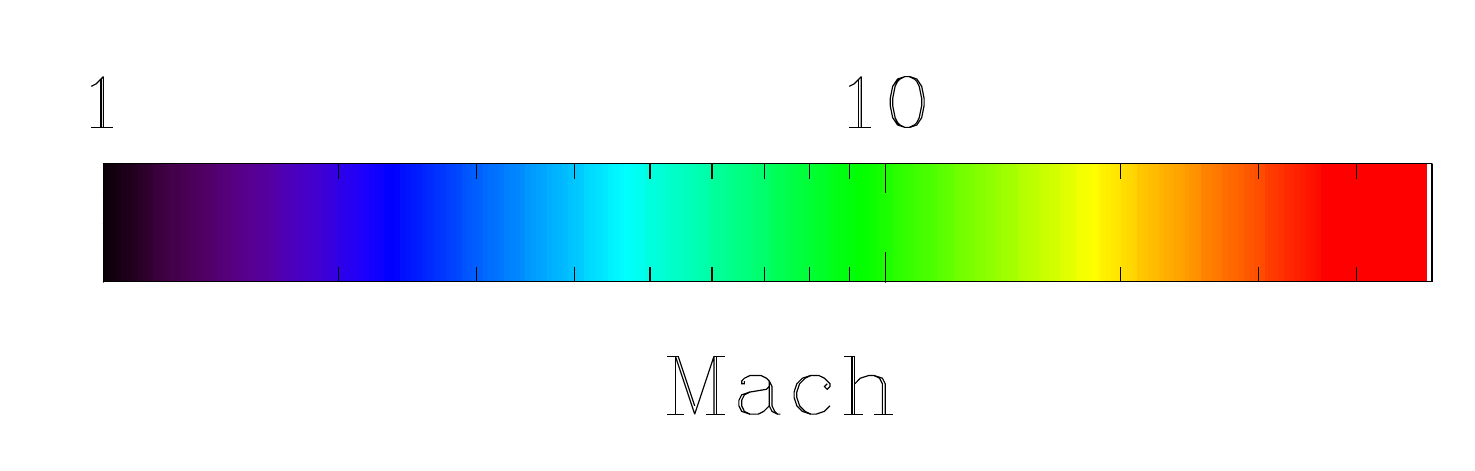}\includegraphics[width=0.5\columnwidth]{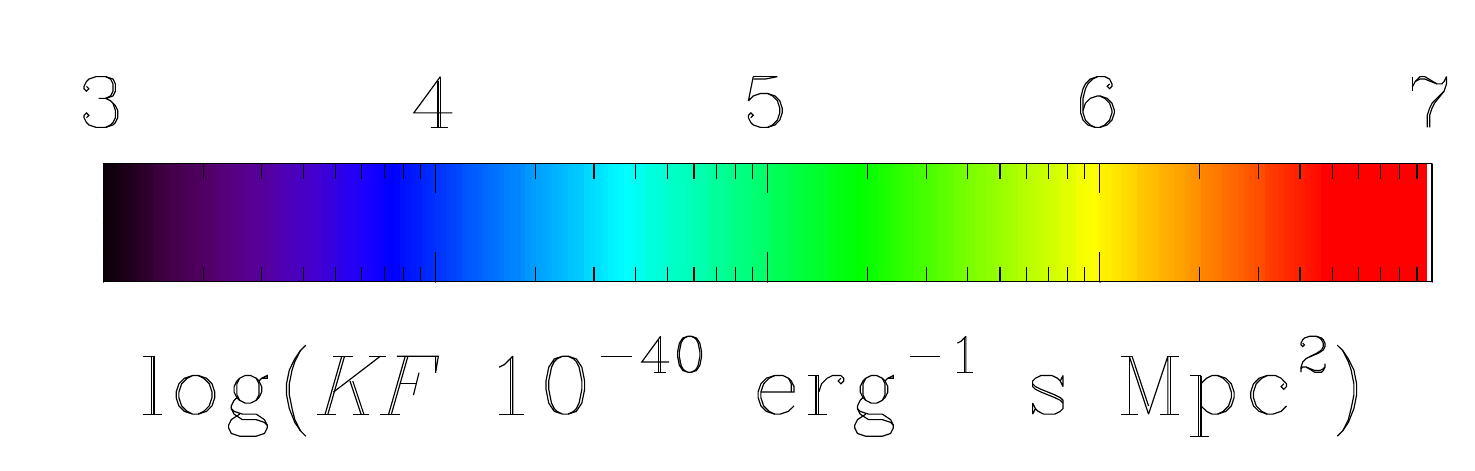}\includegraphics[width=0.5\columnwidth]{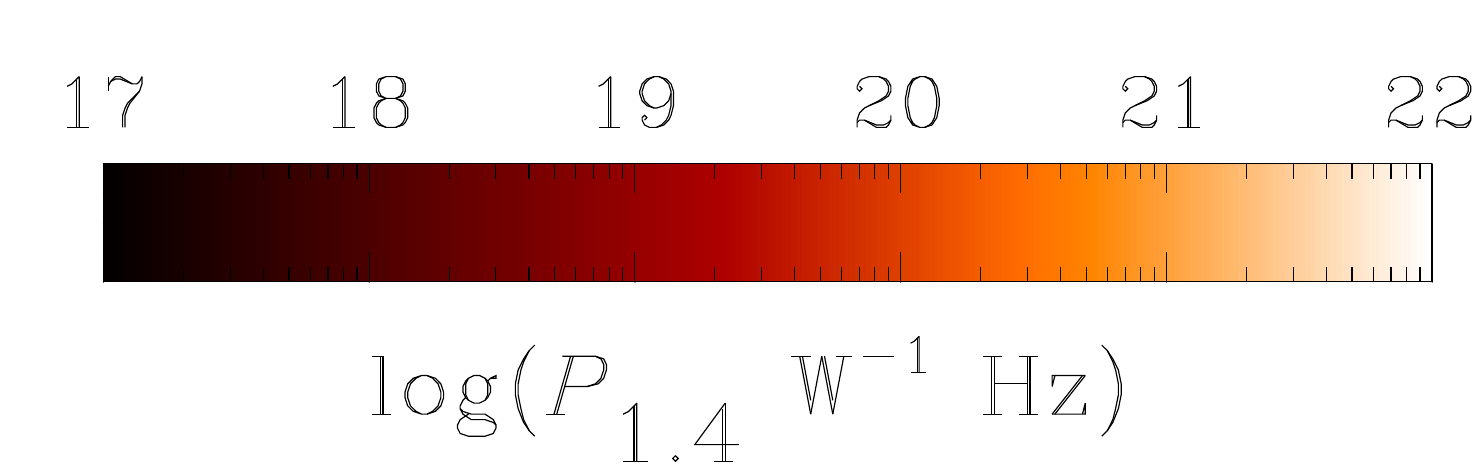}
    \caption{Evolution of gas density (arbitrary units), temperature, Mach number, kinetic energy flux of shocked gas, and radio power at $1.4\,$GHz (from left to right) during a major galaxy cluster binary merger in the sample for two colliding clusters of masses $M_{\rm 200,1}=4.56\times10^{14}\,$M$_{\odot}$ and $M_{\rm 200,2}=2.29\times10^{14}\,$M$_{\odot}$ and impact parameter at $z_{\rm start}$ of $b=140\,$kpc ($b_{200}=0.19$; see text). Each panel shows a region of dimensions $10\times10\times10\,$cMpc$^3$ centred on the main progenitor. Cosmic times $t'$ with respect to the core-passage time $t_{\rm cp}\approx6\,$Gyr ($z_{\rm cp}=0.963$) are shown (from top to bottom): (a) $t'=-609\,$Myr, (b) $t'=-91\,$Myr, (c) $t'=91\,$Myr, (d) $t'=282\,$Myr, (e) $t'=477\,$Myr and (f) $t'=672\,$Myr. Dotted circles in each panel indicate $R_{200,1}(t)$, $R_{{\rm vir},1}(t)$ and $2R_{200,1}(t)$ moving outwards, respectively.}
    \label{fig:merger_evolution_example}
\end{figure*}

\subsection{The non-thermal radio emission model}
\label{sec:radio_emission}

The amount of radio emission produced in structure formation shocks was computed by following the \cite{HB07} model implemented in \cite{Nuza17} \citep[see also][and references therein]{Nuza12}, which assumes that electrons get accelerated at shock fronts according to the DSA mechanism. The total radiation output is regulated by the electron acceleration efficiency, $\xi_{\rm e}$, which fixes the fraction of energy dissipated in shocks channelled into CRe acceleration. The CRes are then assumed to be advected by the downstream plasma and subject to radiative looses and inverse Compton (IC) collisions with cosmic microwave background (CMB) photons. Following~\cite{Kardashev62}, the aged CRe energy spectrum at time, $t$, after injection is
\begin{align*}
    N(\mathcal{E},t) = C_1\,\mathcal{E}^{-s}\left[1-\left(\mathcal{E}_{\rm max}^{-1}+C_2 t\right)\mathcal{E}\right]^{s-2}{\rm ,}
    \nonumber
\end{align*}
\noindent where $C_1$ is a local normalisation constant, $\mathcal{E}$ is the electron energy normalised to its rest mass, $s$ is the slope of the DSA injection spectrum, $\mathcal{E}_{\rm max}$ is the maximum normalised energy of accelerated electrons, and $C_2$ is a constant accounting for synchrotron and IC losses. By integrating the CRe energy spectrum across the shock front it is possible to compute the total non-thermal emission per unit frequency and shock area in the downstream region as a function of magnetic field and local gas properties. 

For the magnetic field, $B$, of every gas particle, we assumed a scaling relation between the field strength and the local gas electron density, $n_{\rm e}$ \citep[see e.g.][]{Dolag01}. Following previous work in \cite{Nuza12}, we adopted
\begin{equation}
    B({\bf r})=B_0\left(\frac{n_{\rm e}({\bf r})}{10^{-4}\,{\rm cm}^{-3}}\right)^\eta,  
\end{equation}

\noindent with $B_0=0.8\,\mu$G and $\eta=0.5$. This parameter choice leads to magnetic field strengths of the order of $\mu$G in galaxy cluster outskirts. For shock fronts propagating from the central region of clusters to their periphery, this implies magnetic field strengths of about $3\,\mu$G down to $0.08\,\mu$G, depending on the densities of the ICM. For instance, for the cluster merger shown in Section~\ref{sec:gischt_lcs}, we obtain magnetic field distributions resulting in $25$, $50$, and $75$ percentile values of $0.76$, $1.32$, and $1.87\,\mu$G at the emission peak, respectively. Although magnetic field strengths obtained in this way are somewhat smaller than those observed at radio-gischt locations, any offset in the resulting radio power can eventually be compensated for by a larger acceleration efficiency. Thus, we consider this approach good enough for our purposes.

The total non-thermal emission behind the shock was then obtained by integrating the electron distribution in the downstream direction, $x$, assuming a constant shock velocity, $v_{\rm sh}=x/t$, which can be related to the sonic Mach number, $\mathcal{M}$. Then, the resulting radio power per unit frequency for gas particle, $i$, reads
\begin{eqnarray}
    P_{\nu,i} 
    & = & 
    6.4 \times 10^{34}\, {\rm erg\,s^{-1}\,Hz^{-1}} \;\;
		\frac{A_i}{{\rm Mpc^2}} \,
		\frac{ n_{{\rm e},i}}{\rm 10^{-4}\,cm^{-3}}\;
		\nonumber                 
		\\
		&& \quad
		\times
		\frac{\xi_{\rm e}}{0.05} \:
		\left(\frac{\nu}{\rm 1.4\,GHz} \right)^{-\frac{s_i}{2}}
		\left(\frac{{T_{{\rm d},i}}}{\rm 7\,keV} \right)^{\frac{3}{2}} \:
		\label{eq:radio_power}     
		\\
		&& \quad
		\times
		\frac{(B_{{\rm d},i}/{\rm \mu G})^{1+\frac{s_i}{2}} }
		{(B_{\rm CMB}/{\rm \mu G})^2 + (B_{{\rm d},i}/{\rm \mu G})^2}
		\;
		\Psi({\cal M}_i)
		{\rm ,}
		\nonumber
\end{eqnarray}

\noindent where $A_i$ represents the surface area associated with the particle, $n_{{\rm e},i}$ is the electron density, $\xi_{\rm e}$ is the electron acceleration efficiency, $s_i$ is the slope of electron energy distribution given by DSA, $T_{{\rm d},i}$ is the post-shock temperature, $B_{{\rm d},i}$ is the post-shock magnetic field, $B_{\rm CMB}$ is the magnetic measure of the CMB energy density, and $\Psi({\cal M}_i)$ is a function that depends on the shock strength. The area corresponding to each SPH particle is proportional to the square of the smoothing length divided by the number of particles within the kernel.

For simplicity, throughout this work we followed \cite{Nuza17} and used $\xi_{\rm e}=5\times10^{-5}$ for the electron acceleration efficiency parameter. The latter was chosen to reproduce the number of clusters hosting radio relics in the  NRAO VLA Sky Survey \citep{Condon98} that have a total flux of $S_{1.4}=100\,$mJy at $1.4\,$GHz for our set of model parameters in the MUSIC-2 simulated galaxy cluster sample \citep[see Section~3.5 of][]{Nuza17}. The resulting radio power outputs were thus taken at face value. Therefore, an extensive modelling of the global radio normalisation is out of the scope of this paper.    

\subsection{The shock detection scheme}
\label{sec:shock_finder}

Shocks were identified in post-processing in the same way as in \cite{Nuza17} (see also \citealt{Nuza12} and references therein). For every gas particle, we defined the shock normal, $\vec{n}\equiv-\nabla P/|\nabla P|$, by evaluating its pressure gradient and later impose a series of conditions to search for true shocks; that is, the shocked region must be in a convergent flow and thermodynamic properties such as gas density and entropy are required to increase when going from the upstream to the downstream region. To estimate the Mach numbers, we used the Rankine-Hugoniot equations for hydrodynamical shocks \citep[see e.g.][]{Landau59} for each of the imposed conditions and take the smaller value as a conservative estimate. We refer the reader to the papers mentioned above for further details. 

\begin{figure}
    \hspace{-0.4cm}
    \includegraphics[width=0.99\columnwidth]{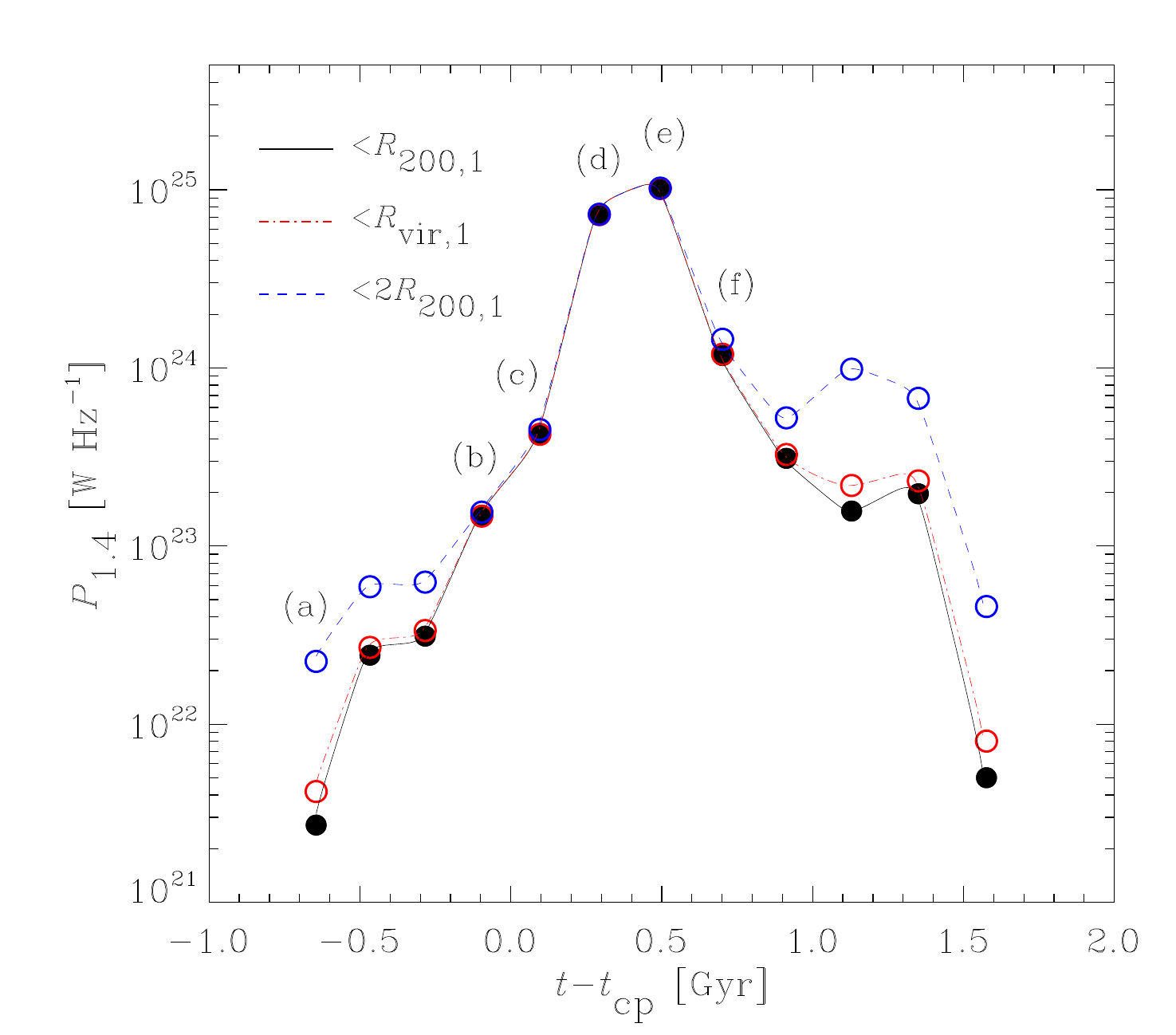}
    \caption{Radio luminosity evolution at $1.4\,$GHz with respect to the core-passage time during the cluster merger event of Fig.~\ref{fig:merger_evolution_example} computed within spherical regions of radii $R_{200,1}(t)$ (solid line), $R_{{\rm vir},1}(t)$ (dash-dotted line), and $2R_{200,1}(t)$ (dashed line) corresponding to the dotted circles shown in the last column. Similarly, letters indicate the different evolutionary stages shown in the previous figure.}
    \label{fig:lc_fig2}
\end{figure}

With this procedure, we estimated shock Mach numbers for the gas component in 279 out of the 324 simulated galaxy cluster regions in \textsc{The Three Hundred} sample, leaving out of the analysis only part of the
less massive groups. Since in this work we are mainly focused on the most massive galaxy groups, which are responsible for the formation of large-scale shocks when interacting, this does not affect our conclusions.

\begin{figure*}
    \centering
    \includegraphics[width=1.04\columnwidth]{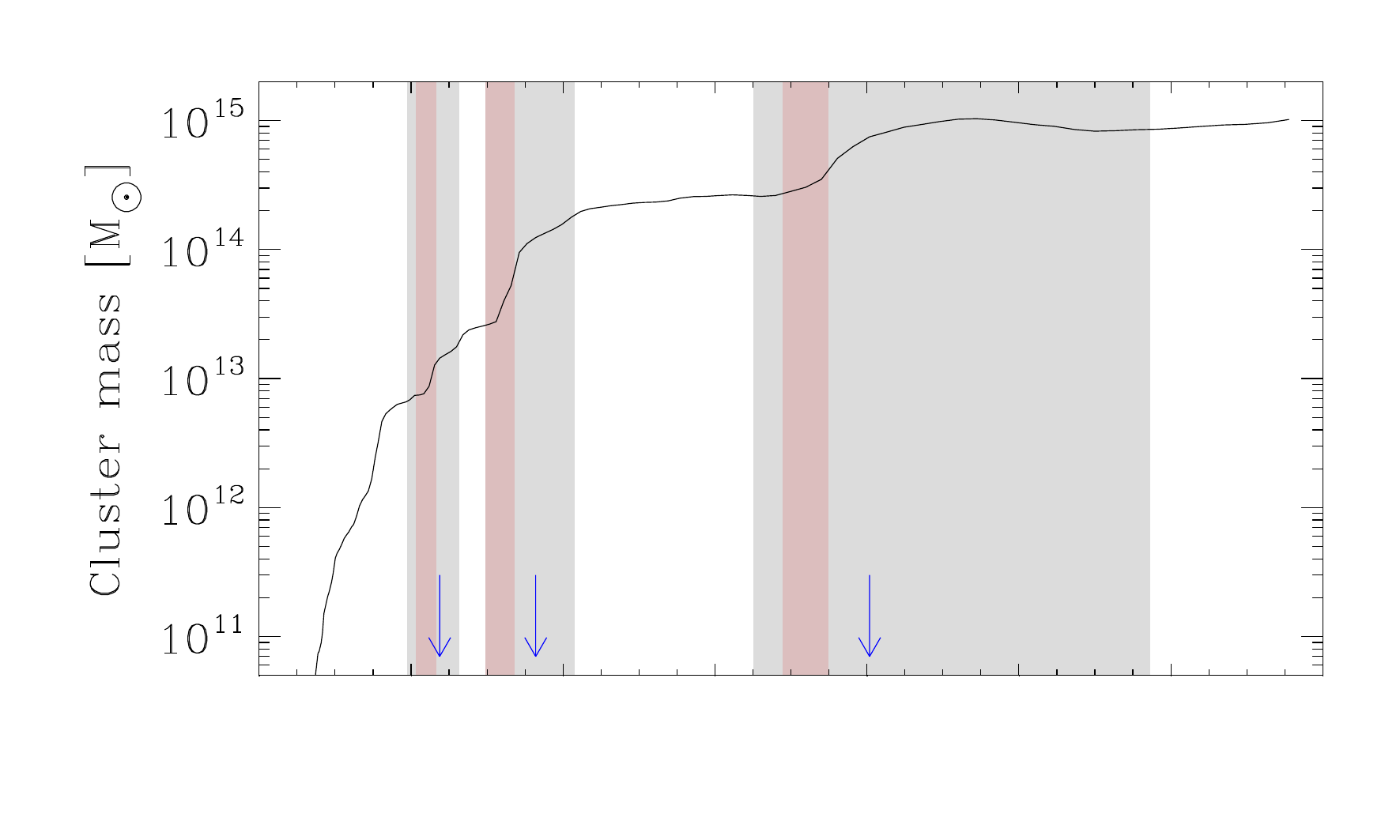}\hspace{-0.5cm}\includegraphics[width=1.04\columnwidth]{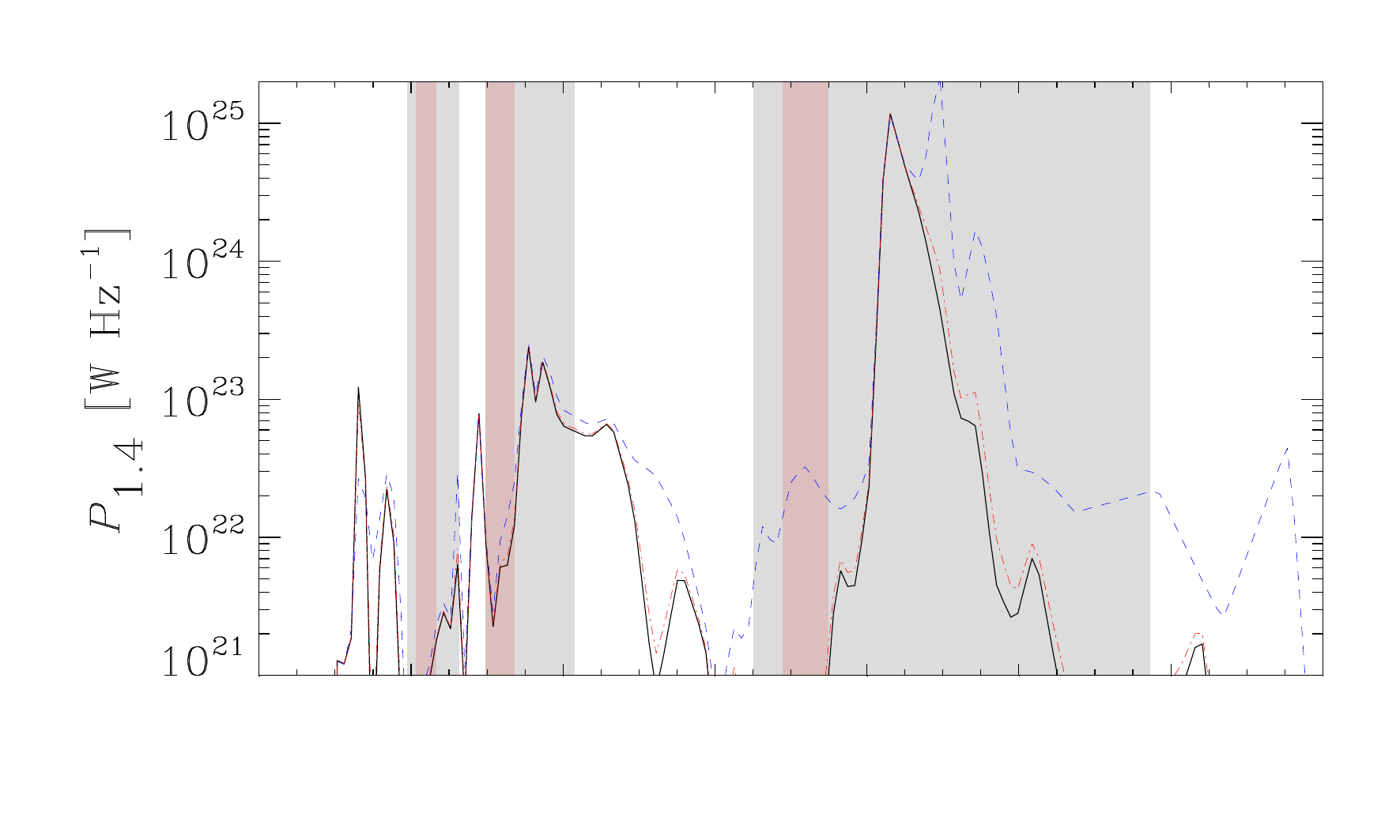}\vspace{-1.4cm}
    \includegraphics[width=1.04\columnwidth]{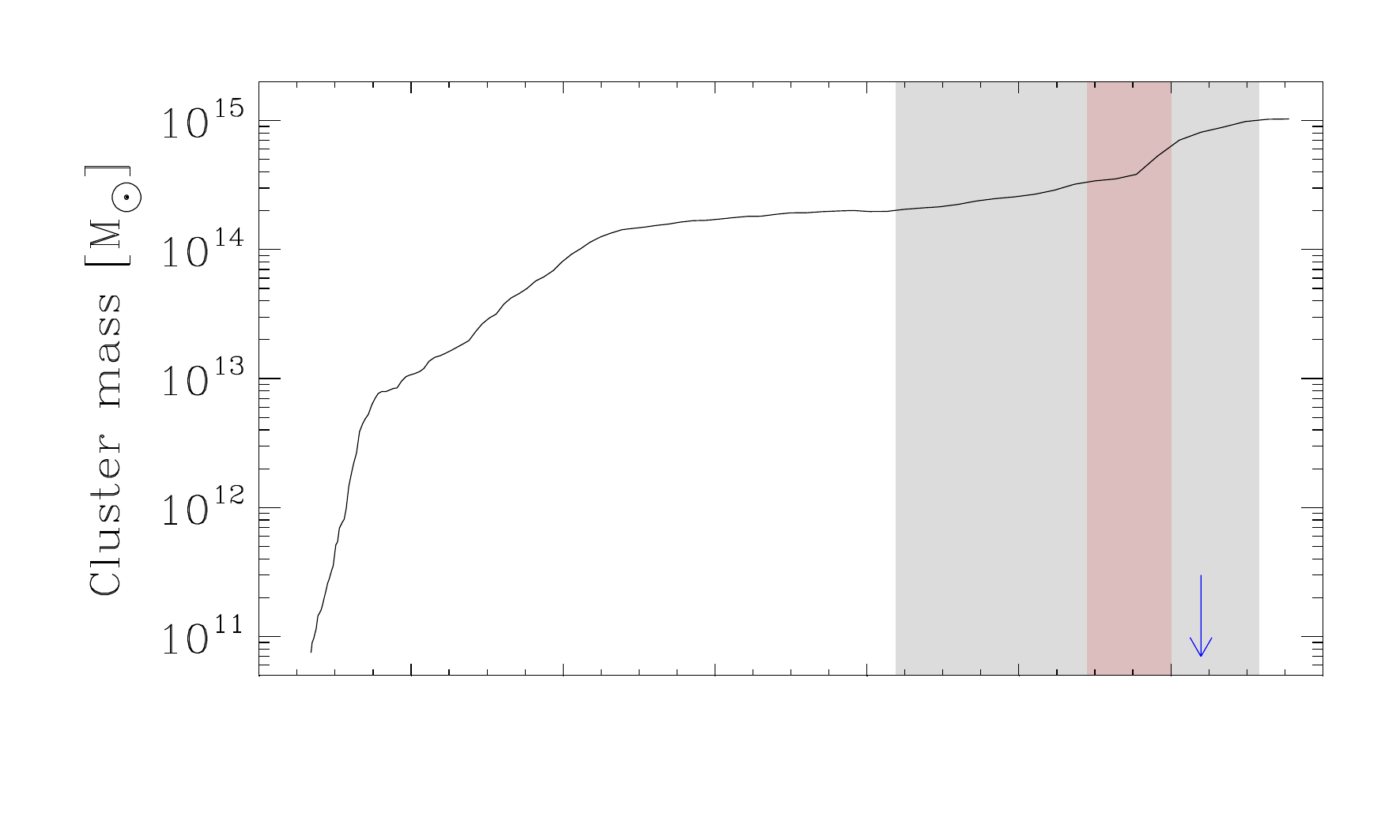}\hspace{-0.5cm}\includegraphics[width=1.04\columnwidth]{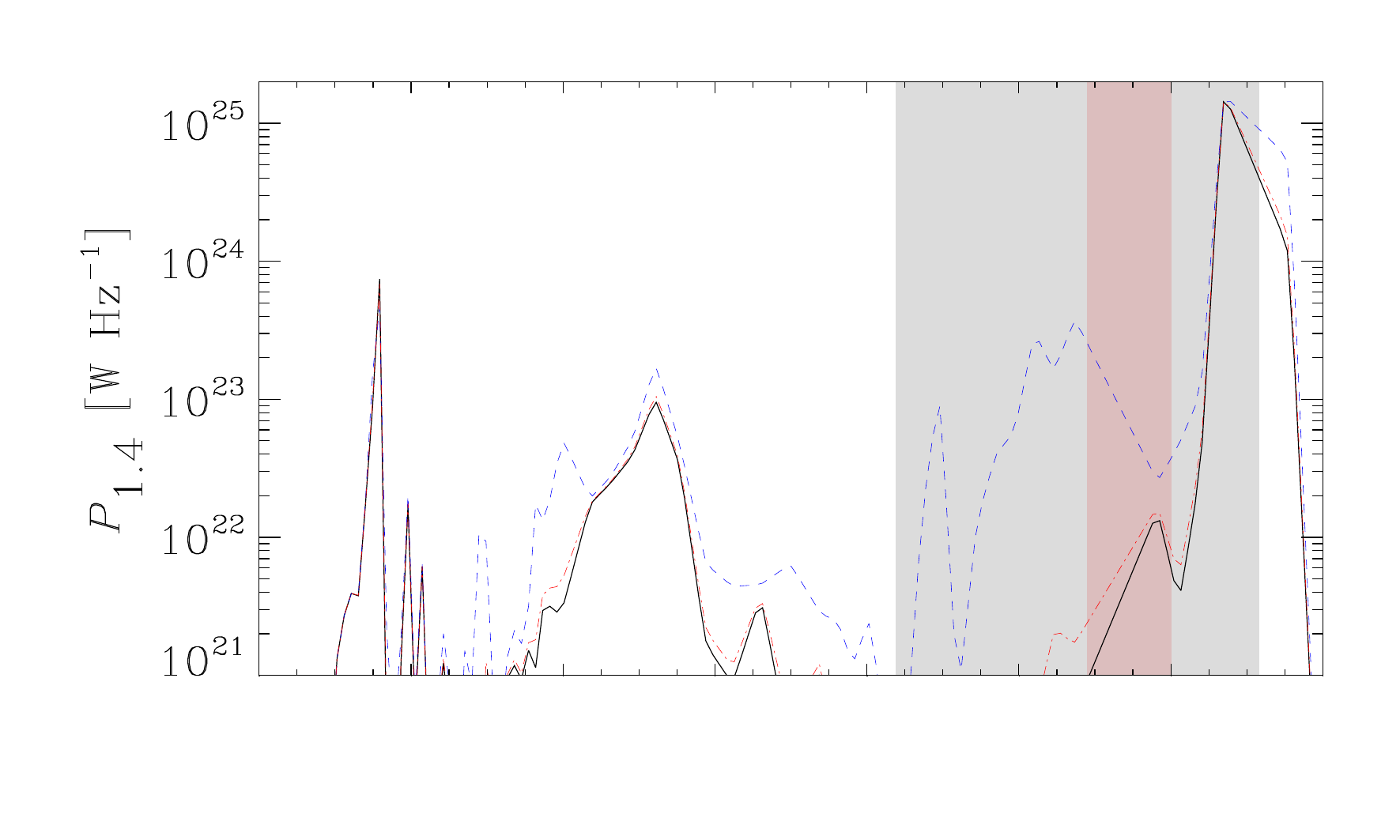}\vspace{-1.4cm}
    \includegraphics[width=1.04\columnwidth]{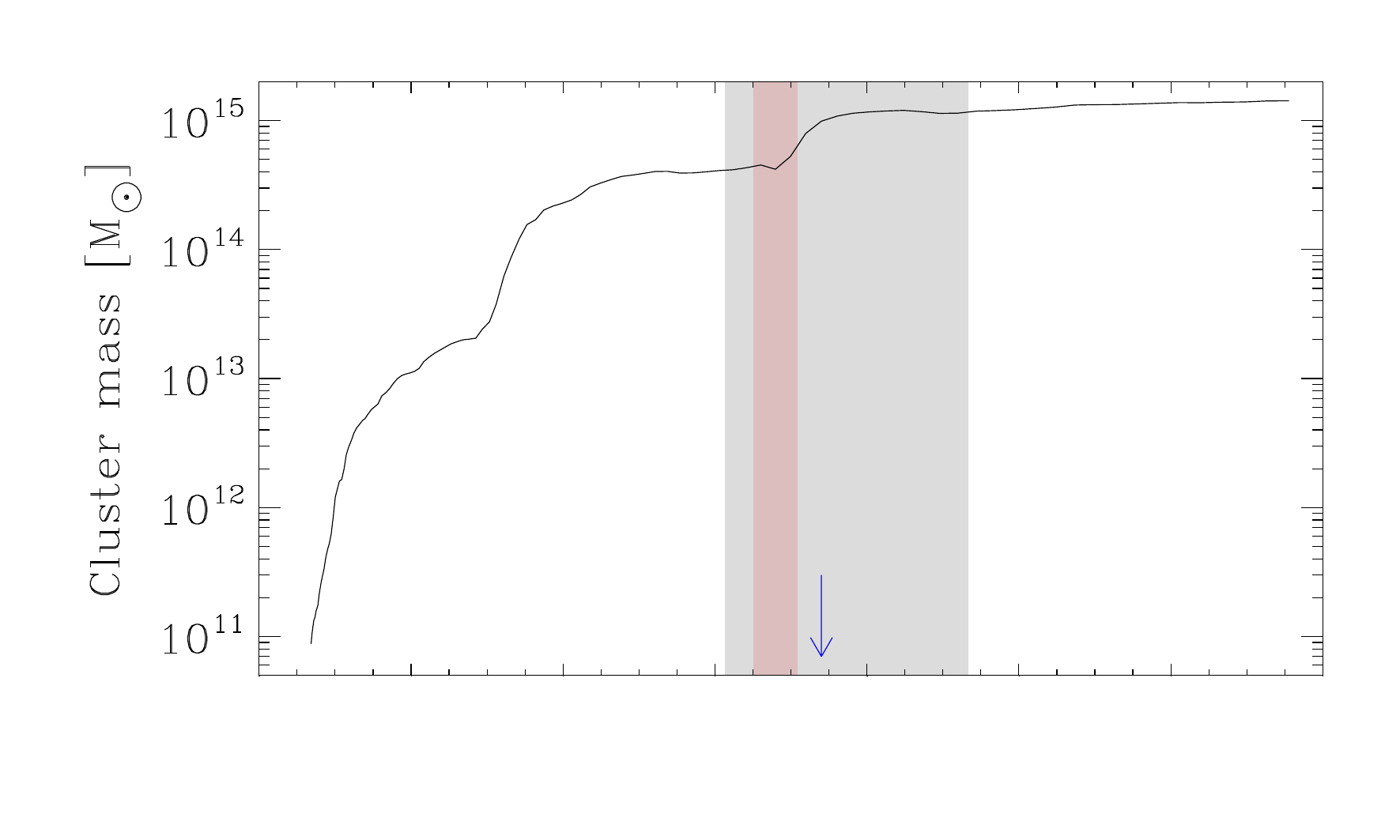}\hspace{-0.5cm}\includegraphics[width=1.04\columnwidth]{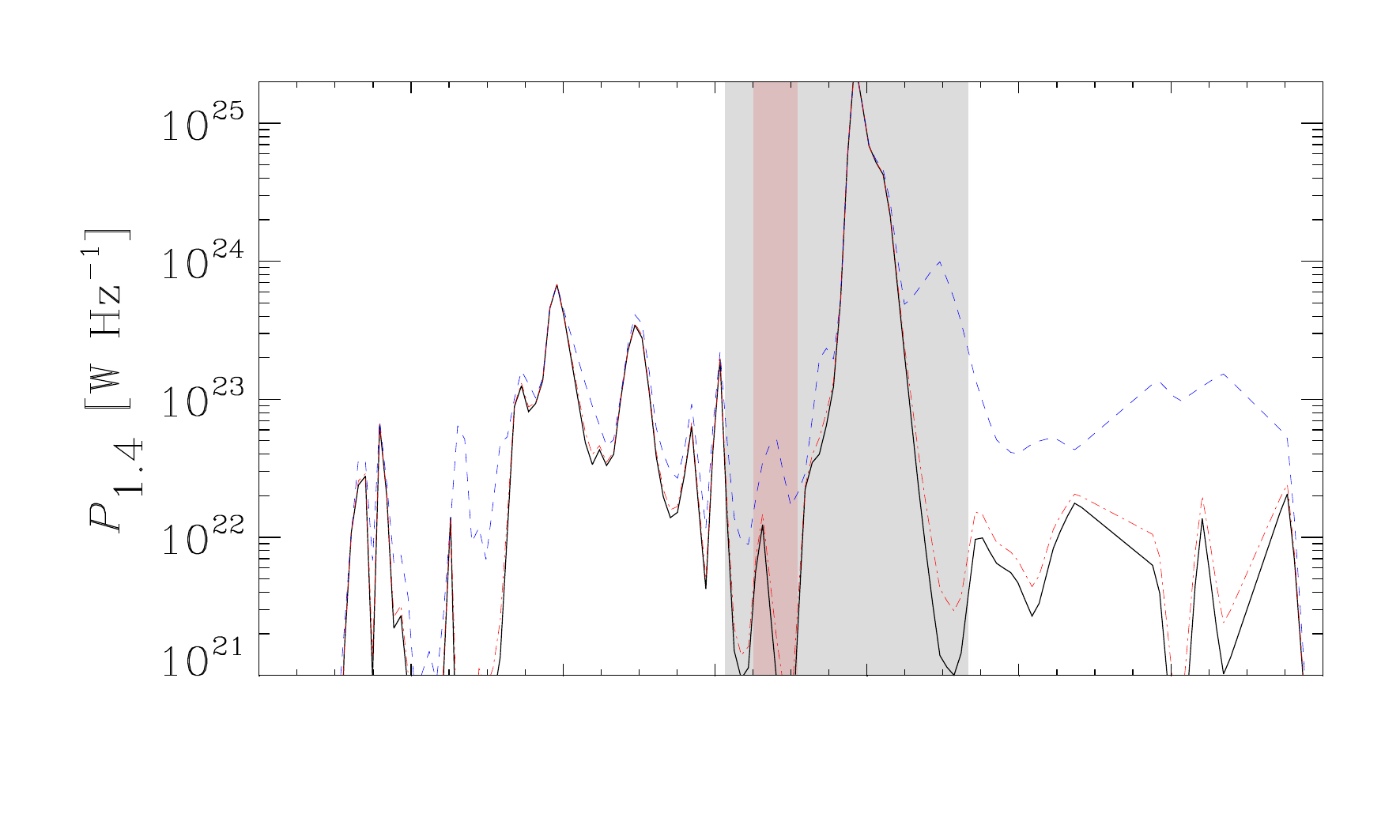}\vspace{-1.4cm}
    \includegraphics[width=1.04\columnwidth]{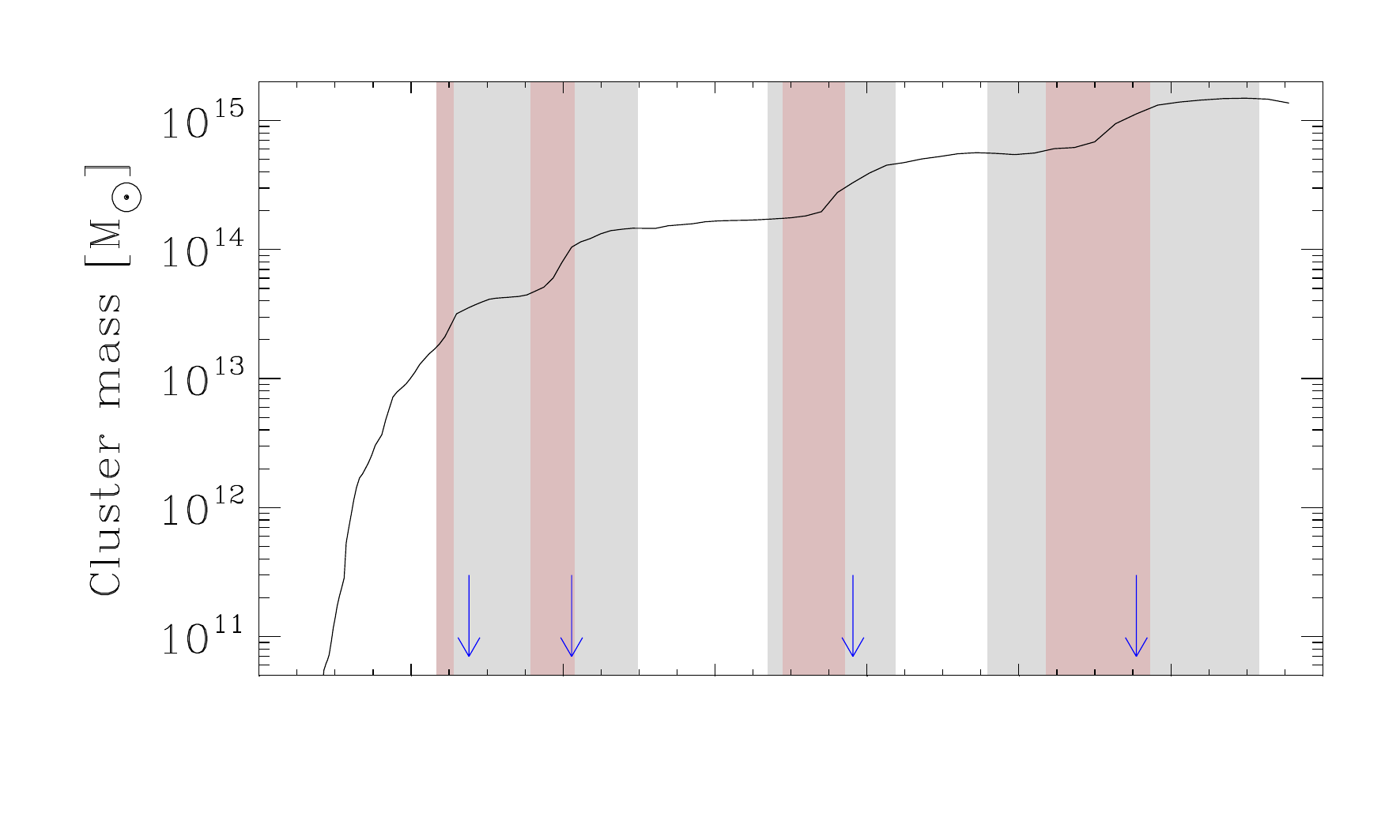}\hspace{-0.5cm}\includegraphics[width=1.04\columnwidth]{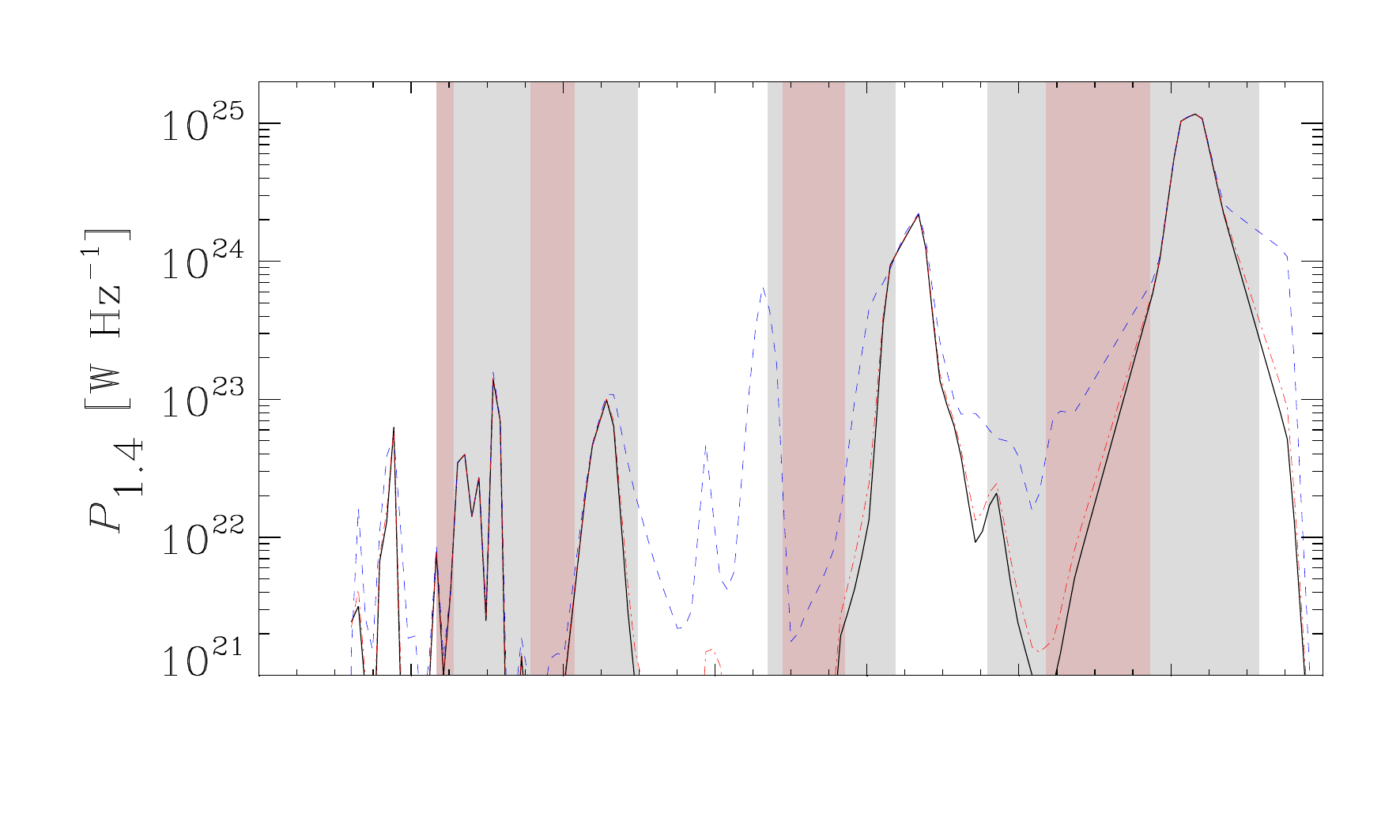}\vspace{-1.4cm}
    \hspace*{0.09cm}\includegraphics[width=1.04\columnwidth]{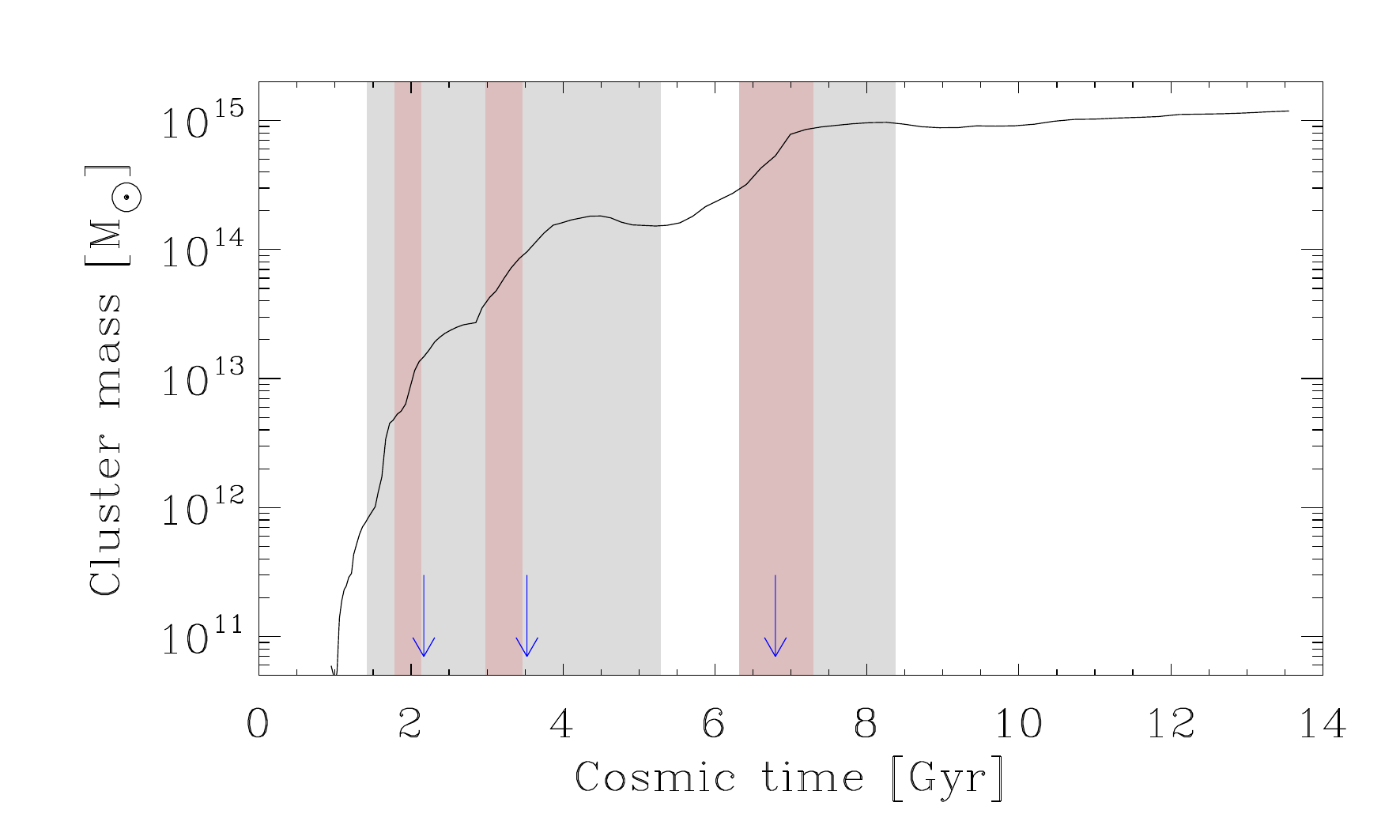}\hspace{-0.5cm}\includegraphics[width=1.04\columnwidth]{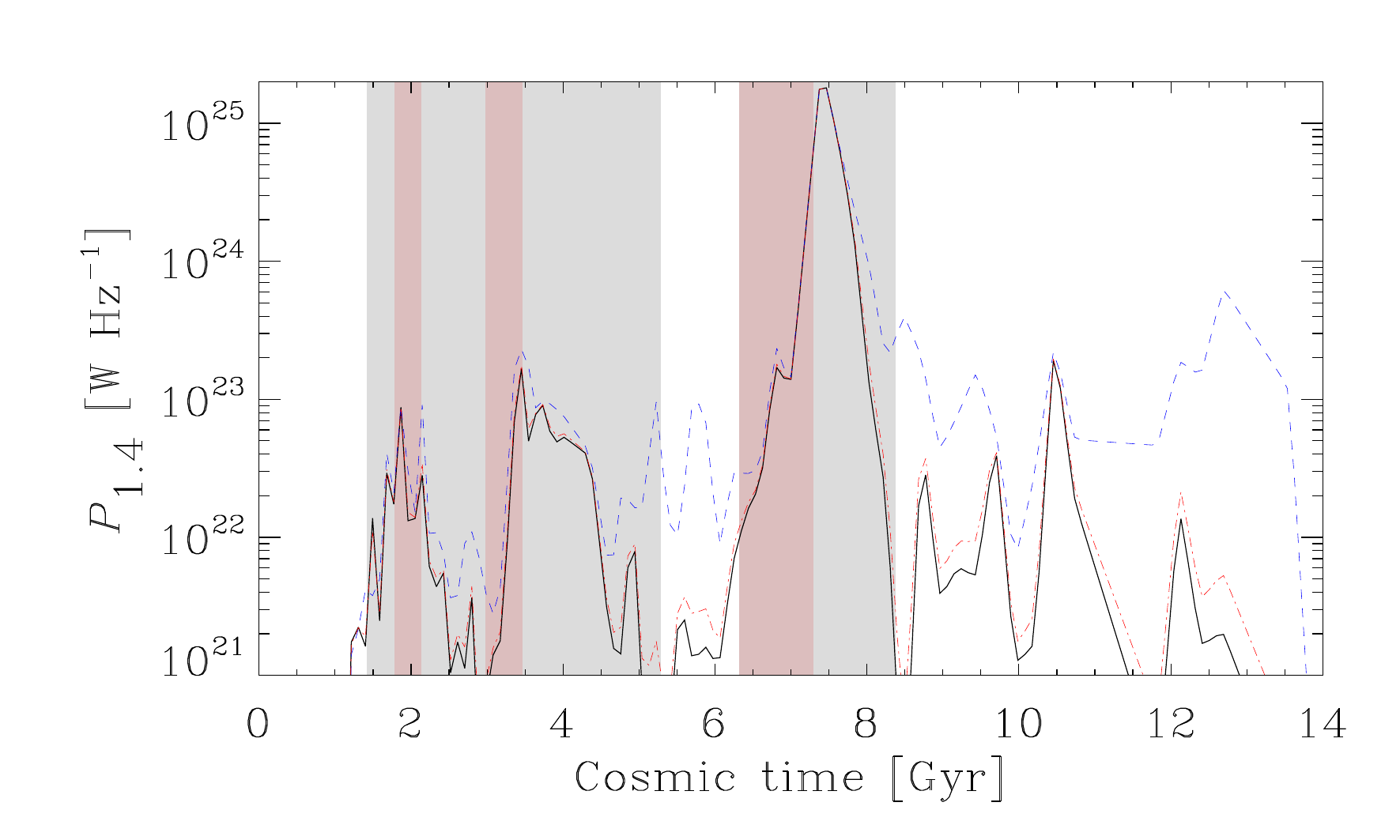}
    \caption{Mass assembly history of the main galaxy cluster characterised by $M_{200,1}(t)$ (left-hand panels) and corresponding radio power per unit frequency at $1.4\,$GHz (right-hand panels) as a function of cosmic time for different cluster regions in our sample. Solid, dash-dotted, and dashed lines show all radio luminosity measured within $R_{200,1}(t)$, $R_{\rm vir,1}(t)$, and $2R_{200,1}(t)$, respectively. Major mergers with $\Delta M/M \geq 0.5$ are highlighted by the vertical grey and light red colour bands corresponding to the $t_{\rm after}-t_{\rm before}$ and $t_{\rm end}-t_{\rm start}$ time intervals. Blue arrows in the left panels indicate the core-passage time of each merger.}
    \label{fig:radioLCs}
\end{figure*}

\section{Results}
\label{sec:results}

\subsection{Galaxy cluster merger rate in the sample}
\label{sec:merger_rate}

Mergers are transient phenomena in the life of galaxy groups and clusters. Within a hierarchical cosmological scenario it is generally expected that the most massive mergers are less frequent in comparison to their lower-mass counterparts \citep[e.g.][]{Genel10}.

To quantify the occurrence of collisions in our cluster merger sample (see Section~\ref{sec:merger_sample}), we counted the number of mergers per redshift bin for all simulated cluster regions and divided by the corresponding cosmic time interval. As a proxy of merger mass, we used the main progenitor mass, $M_{{200},1}$, defined at $t_{\rm start}$, together with the merger mass ratio, $\mu$. Fig.~\ref{fig:merger_rate} shows the redshift evolution of the differential number of mergers per unit time normalised by the number of cluster regions in a given primary cluster mass bin. The left-hand panel shows the merger rate for all mergers involving galaxy groups and clusters after imposing the condition $M_{{200},1}\geq 10^{13}\,$M$_{\odot}$. This curve peaks around $z\sim2$ with a median redshift of $z=1.32$, which can be explained by the presence of low-mass systems in the sample. In fact, the latter is dominated by galaxy groups with a median mass of $M_{200,1}=8.76\times10^{13}\,$M$_{\odot}$ for the primary cluster and $M_{200,2}=3.96\times10^{13}\,$M$_{\odot}$ for the secondary. This plot shows the merger rate of the complete sample studied in this work without any cut in merger mass ratio, allowing all cluster regions to be included in the normalisation. The decreasing trend towards low redshifts reflects both the fact that the more abundant, lower-mass systems tend to merge at earlier times, in contrast to the more massive systems, and that the cosmic time intervals decrease with redshift.

\begin{table*}
    \caption{Best-fit parameters of the median radio luminosity evolution at $1.4\,$GHz for the curves shown in Figs.~\ref{fig:mean_radioLCs} (rows 1--4) and~\ref{fig:mean_radioLCs2} (rows 5--8) assuming the fitting formula defined in Eq.~(\ref{Eq:bestfit}). All fits are performed within $R_{200,1}(t)$ except for rows 3 and 4 as indicated.
    }
    \centering
    \begin{tabular}{lcccccc}
    \hline\hline
    Parameters & $a_0$ & $s_0$ & $s_1$ & $\sigma$ & $t_0$ & $\gamma$ \\
    & [$10^{20}\,$W\,Hz$^{-1}\,$Gyr] & [$10^{20}\,$W\,Hz$^{-1}\,$Gyr$^{-1}$] & [$10^{20}\,$W\,Hz$^{-1}$] & [Gyr] & [Gyr] & \\
    \hline
    groups & $174\pm19$ & $-2.2\pm2.4$ & $20.0\pm3.3$ & $0.86\pm0.12$ & $0.03\pm0.04$ & $6.8\pm2.5$\\ 
    clusters & $6020\pm250$ & $49.0\pm9.4$ & $91.8\pm8.9$ & $0.42\pm0.02$ & $0.39\pm0.03$ & $1.76\pm0.27$\\
    groups ($\leq 2R_{200,1}$) & $279\pm82$ & $18\pm14$ & $124\pm12$ & $1.00\pm0.87$ & $0.16\pm0.06$ & $5.9\pm3.6$\\ 
    clusters ($\leq 2R_{200,1}$) & $9820\pm480$ & $370\pm110$ & $665\pm53$ & $0.68\pm0.05$ & $0.36\pm0.02$ & $3.34\pm0.52$\\
    \hline
    early  & $188 \pm 16 $ & $ 10 \pm 3 $ & $30.3 \pm 2.9$ & $0.75 \pm 0.07$ & $ 0.03 \pm 0.02 $ & $6.8\pm1.6$\\
    late & $ 8490 \pm 320 $ & $ 29.5 \pm 7.6 $ & $ 79.0 \pm 8.2 $ & $ 0.42 \pm 0.02 $ & $ 0.51 \pm 0.03 $ & $1.56\pm0.22$\\
    early ($b<200\,$kpc) & $133 \pm 14$ & $-2.5 \pm 1.4$ & $10.8 \pm 2.4$ & $1.00 \pm 0.04$ & $-0.02 \pm 0.04$ & $8.0\pm 2.5$\\
    late ($b<200\,$kpc) & $7361 \pm 780$ & $-1.0 \pm 3.3$ & $15.9 \pm 5.7$ & $0.38 \pm 0.03$ & $0.25 \pm 0.03$ & $2.89 \pm 0.66$\\
    \hline\hline
    \end{tabular}
    \label{tab:bestfit_param}
\end{table*}

In the right-hand panel of Fig.~\ref{fig:merger_rate}, the merger rates are split in two different cluster mass bins of the main progenitor corresponding to galaxy groups or less massive clusters with $M_{\rm 200,1}$ between $10^{13}$ and $10^{14}\,$M$_{\odot}$ (the `group' sample), and more massive clusters with $M_{\rm 200,1}\geq10^{14}\,$M$_{\odot}$ (the `cluster' sample). For each mass bin, we performed an additional constraint in the merger mass ratio for $\mu>0.5$ (solid line) and $\mu>0.75$ (dashed line) that will decrease the number of selected cluster regions with each additional cut. As a consequence, these cuts in the merger mass ratio will also result in the selection of quite massive secondary clusters. For instance, in the $\mu>0.5$ case, the median galaxy group mass within $R_{200}$ in the less massive bin (solid blue line) is $2.84\times10^{13}\,$M$_{\odot}$ for the primary cluster and $2.14\times10^{13}\,$M$_{\odot}$ for the secondary. Similarly, in the more massive bin (solid red line), the corresponding galaxy cluster masses are $2.88\times10^{14}$ and $1.96\times10^{14}\,$M$_{\odot}$ for the primary and secondary clusters, respectively.

As has already been shown in the plot, these curves show that, on average, less massive systems tend to merge early in the evolution of the Universe, reaching maximum merger rates at $z\sim2$. This is in contrast to the more massive cluster mergers that preferentially peak around $z\sim0.5$ and that are responsible for most of the observed radio relics, which have a mean redshift distribution of $z\sim0.3$ \citep[e.g.][]{Jones23}. In general, when increasing the merger ratio, rates tend to shift towards lower redshifts. When comparing mergers of different cluster masses, it is evident that less massive mergers take place more often in comparison to more massive ones, although some overlap is possible both in merger rate and time, inverting the trends. These facts are consistent with the hierarchical nature of structure formation whereby more massive systems build up from smaller ones as cosmic time elapses. Interestingly, this plot can be used to estimate the mean number of mergers that a cluster of a given mass in our sample will experience during its lifetime. For instance, clusters in the most massive bin with main progenitors of about $2-3\times10^{14}\,$M$_{\odot}$ will experience, on average, at least one major merger at redshifts less than $z\sim1$. 

We stress, however, that by construction our merger catalogue is biased towards massive galaxy cluster regions at $z=0$. Therefore, the merger rate of groups in our sample inhabiting protocluster regions does not fully represent that of the general population of less massive objects at higher redshifts. Nonetheless, here we are mainly interested in characterising the mass assembly evolution of the selected galaxy cluster regions in our catalogue. We leave for future work a more detailed assessment of the merger rate evolution in galaxy groups and clusters located in different environments and its relation with objects in a full cosmological volume.

\subsection{Gischt radio light curves}
\label{sec:gischt_lcs}

To illustrate the evolution of merger shocks during galaxy cluster collisions in a cosmological context, in Fig.~\ref{fig:merger_evolution_example} we plot the gas density, temperature, Mach number, kinetic energy flux of shocked gas, and radio luminosity at different times with respect to the core-passage for a pair of colliding clusters of masses $M_{\rm 200,1}=4.56\times10^{14}\,$M$_{\odot}$, $M_{\rm 200,2}=2.29\times10^{14}\,$M$_{\odot}$, and an `impact parameter'\footnote{This parameter gives an estimate of the extrapolated distance between the colliding clusters at core-passage (see Section~\ref{sec:impact_param}).} evaluated at $t_{\rm start}$ of $b=140\,$kpc ($b_{200}=0.19$ in $R_{200,1}$ units) that happens to almost merge in the plane of the figure. The low impact parameter suggests that this particular system is undergoing a fairly radial collision, as will be discussed in the next section. The kinetic energy flux of shocks, on the other hand, is defined as $0.5\rho_{\rm g}v_{\rm sh}^3$, where $\rho_{\rm g}$ is the gas density in the downstream region and $v_{\rm sh}$ is the shock velocity. In the last column, the evolution of $R_{200,1}(t)$, the virial radius $R_{{\rm vir},1}(t)$, and $2 R_{200,1}(t)$ in comoving coordinates are also shown in each panel.

According to our radio emission model, we expect that radio luminosity is maximised in regions with higher density, temperature, and Mach number, although the Mach number dependence is weak for $\mathcal{M}\gtrsim2$. This information is encapsulated in the so-called kinetic energy flux of the shock fronts. Radio emission is preferentially produced in shocked regions with higher kinetic energy fluxes because this quantity is proportional to gas density and also depends strongly on the shock velocity, which correlates with downstream temperature. The correlation between the kinetic flux and the non-thermal radio output is evident in the case of merger shocks formed after core-passage. However, despite their higher Mach numbers, external shocks do not show appreciable radio emission owing to the lower gas densities. Conversely, as is shown before the core-passage (rows a and b), the relatively high kinetic flux seen in the vicinity of clusters (boosted by the higher gas densities close to the central cluster regions) for Mach numbers below $\mathcal{M}\lesssim 2$ does not correlate with new significant radio emission outputs.

Before the core-passage (rows a and b), gas temperature rises between the two merging systems as they approach each other. Axial shocks (i.e. those that roughly propagate in the direction of the merger axis) are launched right before the core-passage, in agreement with the work of \cite{Ha18}. Therefore, as can already be seen in the last panel of row b, the radio luminosity output within $R_{200,1}$ at core-passage is expected to be different from zero even at this early stage. The presence of external shocks are mainly seen in the cluster outskirts as a result of gas accretion towards the central cluster. This is a general feature of galaxy clusters, irrespective of their dynamical state, as it is expected that shocks are developed when accreting gas falls onto the potential well of high-density regions \cite[see ][for an analysis of shocks around filaments and clusters in \textsc{The Three Hundred} sample]{Rost24}. Additionally, some minor internal shocks are produced by the merging of smaller sub-structures inside the cluster region.

After the core-passage (rows c, d, e, and f), the formation of axial shocks is clearly seen. Typically, these shocks have low Mach numbers of $\mathcal{M}\sim2-5$, in contrast to external shocks, which, given the smaller ICM densities, tend to have larger Mach numbers. Despite the low Mach number values, these shocks produce a significant heating of the ICM during their existence. As time elapses, merger shocks expand towards the external cluster regions displaying  synchrotron emission that is mainly produced by CRe cooling downstream, forming the so-called radio relic structures or gischts. In this particular example, the radio relic emission peaks at $\sim500\,$Myr after the core-passage (row e), as will be shown below.

The radio luminosity evolution of this cluster merger with respect to the core-passage time for all radiation measured within spheres of radii $R_{200,1}(t)$, $R_{{\rm vir},1}(t)$ and $2R_{200,1}(t)$ centred on the main subcluster can be seen in Fig.~\ref{fig:lc_fig2} (solid, dash-dotted, and dashed lines, respectively). We note that, in this work, we have not removed any spurious radiation eventually produced by AGN shocks generated as a result of the implemented black hole growth in the simulations, although from visual inspection of the merger snapshots this possible source of contamination does not seem to play a significant role in relation to the relic emission. The first two distance scales mentioned above are representative of the typical size of the main cluster halo. The last scale (i.e. twice the $R_{200,1}$ distance) is also  considered to include radiation eventually escaping from the surveyed volume as radio shocks expand to cluster outskirts at the expense of including some spurious emission produced by other, non-related, shocks in the ICM.

As can be seen in Fig.~\ref{fig:lc_fig2}, the luminosity peak is well captured inside $R_{200,1}(t)$ (see point (e) marked in the plot), being essentially the same as the dashed line at $t_{\rm peak}$. Before this time, the difference between the solid and dashed lines is mainly due to noise from other sources because the $R_{200,1}(t)$ volume is already large enough to contain all relic emission from its onset. After the maximum luminosity is reached, the dashed line decays slower than the solid line because, in this case, more radiation ends up within the search volume. Specifically, the increase in the amount of radiation compared to the $R_{200,1}(t)$ sphere between the radio peak at $\sim500\,$Myr and $\sim1.5\,$Gyr is about $\sim40\%$. One has to keep in mind, however, that a significant fraction of this radiation comes from unwanted sources such as external shocks and/or from other smaller accreting structures (see Fig.~\ref{fig:merger_evolution_example}).  Any additional filtering of spurious radiation not linked to the expanding merger shocks is out of the scope of this work and it will be considered in forthcoming papers. Finally, the emission measured within the virial radius and $R_{200,1}(t)$ is essentially the same at these redshifts. At late times, the difference between $R_{200,1}$ and the virial radius is typically of the order of $30\%$ for most of the cluster progenitors studied here, boosting the search volume roughly by a factor of two, but the resulting light curves are not dramatically affected, especially for the most luminous merging systems.

\subsection{Cluster merger radio shocks across cosmic history}

The evolution of mass and radio power during galaxy cluster mergers over the entire cluster lifetimes can be better seen in Fig.~\ref{fig:radioLCs} for several cluster examples in our sample. The left-hand panels show $M_{200,1}(t)$; that is, the mass accretion history of the main halo in the re-simulated cluster regions as a function of cosmic time, $t$. The shaded regions span the time intervals $t\in[t_{\rm start},t_{\rm end}]$ and $[t_{\rm before},t_{\rm after}]$ (light red and grey colour bands, respectively) during cluster interactions, being $t_i$ the different characteristic times defined in Section~\ref{sec:merger_sample}. The mass accretion history of each system is reconstructed from the AHF merger trees after identifying the galaxy cluster progenitor at each cosmic time with the aid of a merit function that depends on the number of particles in each parent halo \citep[see e.g.][]{Mostoghiu19}. As already mentioned, in this work, we are mainly focused on major mergers where the mass increase of the main progenitor is at least $50\%$ \citep[see][]{Contreras22}. During these events there is an evident increase in the cluster mass after $z_{\rm start}$, being generally flat or moderate after $z_{\rm end}$. For every major merger in our catalogue we also estimate the core-passage time of the two most massive haloes by computing their mutual distance during the merger interval to determine their minimum relative distance and the change of direction in the associated relative velocity vector. Blue arrows in Fig.~\ref{fig:radioLCs} indicate the core-passage time of each merger in our galaxy cluster examples, suggesting that, in most cases, the core-passage time is similar to $t_{\rm end}$. The latter has been visually confirmed by inspecting the evolution of individual merger snapshots in the sample. As is seen in the plots the evolution of clusters indicate that only a few major mergers take place during a given galaxy cluster's lifetime. 

\begin{figure}
    \hspace{-0.4cm}
    \includegraphics[width=0.99\columnwidth]{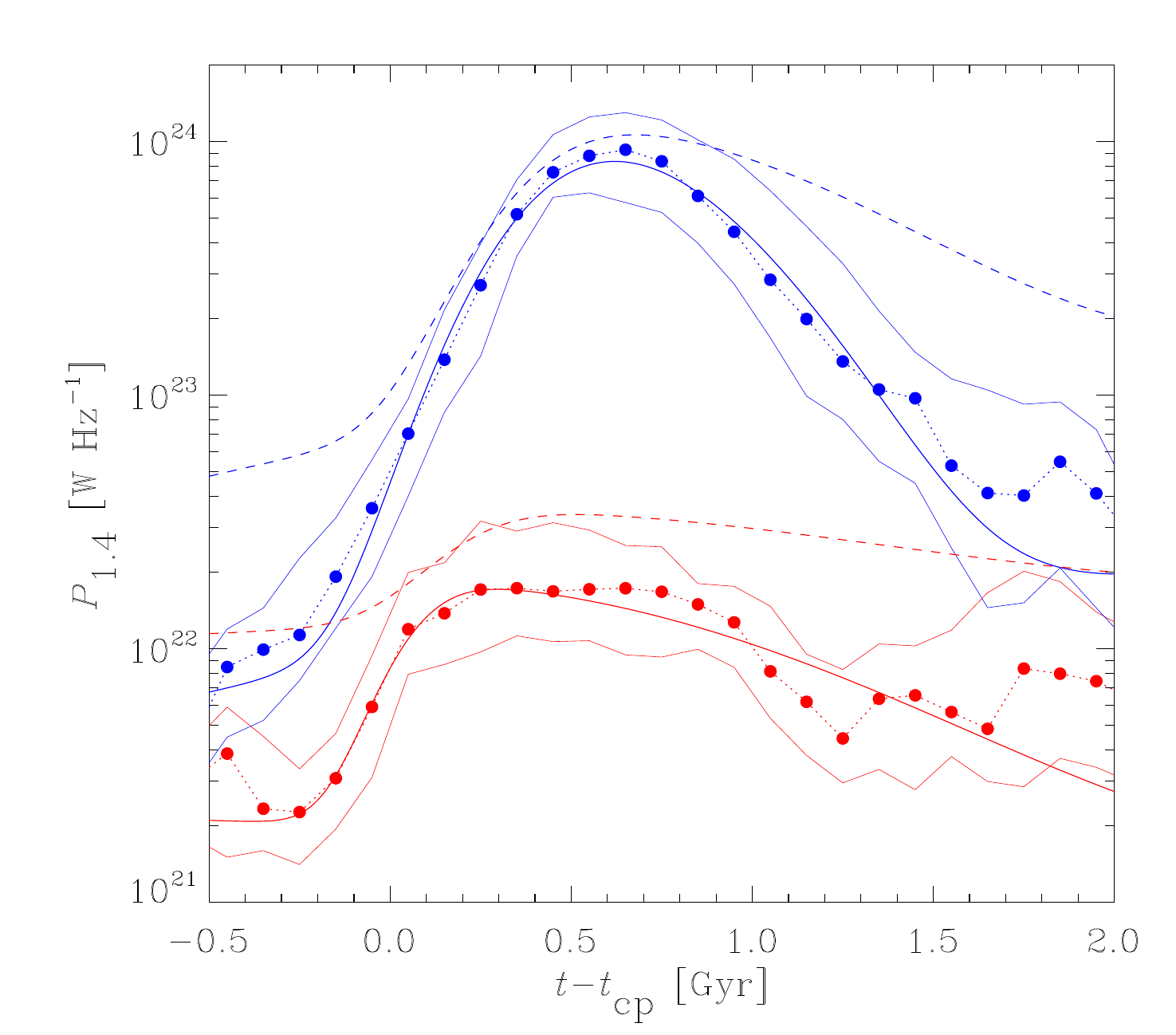}
    \caption{Median light curves at $1.4\,$GHz within $R_{200,1}(t)$ during cluster merger events for galaxy groups with $10^{13}\leq M_{\rm 200,1}/{\rm M}_{\odot}<10^{14}$ (solid red circles) and clusters with $M_{200,1}\geq10^{14}\,$M$_{\odot}$ (solid blue circles) as a function of time with respect to the core-passage. In each case, thin solid lines correspond to a decile above and below the median value. Best-fitting functions within $R_{200,1}(t)$ and $2R_{200,1}(t)$ are also shown as solid and dashed lines, respectively.}
    \label{fig:mean_radioLCs}
\end{figure}

The solid lines in the right-hand panels of Fig.~\ref{fig:radioLCs} show the corresponding radio luminosity light curves at $1.4\,$GHz for radiation produced within $R_{200,1}(t)$ by CRe in all shocks within the surveyed volume. As was mentioned before, this distance scale will mainly pick emission from inner (merger) shocks in the main cluster halo produced during galaxy cluster collisions as they evolve from onset until they reach the external regions of the ICM. The radiation present inside the virial radius of each main progenitor is also shown (dash-dotted lines). A larger radius will also include radiation coming from outer (accretion) shocks around galaxy clusters and/or surrounding filaments, an outcome that we would like to avoid. Nevertheless, for completeness, and following the same spirit of Fig.~\ref{fig:lc_fig2}, we also plot the evolution of the emission measured within the $2R_{200,1}(t)$ spheres using dashed lines. These figures indicate that, according to our model, every merger event is followed by a peak of radio emission produced as a result of the formation of merger shocks and the consequent CRe acceleration in the presence of magnetic fields. This is especially evident after the core-passage of the secondary cluster (see arrows). Generally speaking, the position and magnitude of the radio emission peaks are well captured within $R_{200,1}(t)$ as demonstrated by the relative invariance of most of the maximums in the $2R_{200,1}(t)$ case. However, after the curves have peaked, more radiation is included in the latter as a consequence of the larger volume, which increases by a factor of 8 relative to that of the search radius of solid lines.

These radio light curves represent the actual evolution of gischt-related emission in our simulated galaxy clusters as the shock identification and radio luminosity model have been run for all available snapshots, independent of the cluster merger catalogue presented in Section~\ref{sec:merger_sample}. In fact, other radio peaks can be seen within $R_{200,1}$ at cosmic times with no major mergers, although at much lower luminosities. This demonstrates that, in fact, observed radio relics represent only the tip of the iceberg of the total non-thermal radiation produced during structure formation. In particular, at very early times, even during major mergers (see shaded regions), peak luminosities are always lower than their late-time counterparts, where the most massive mergers occur. This is consistent with the fact that most relic detections available today correspond to moderate or low redshifts; that is, $z\lesssim0.5$.

An interesting feature that can be observed in some of the examples is the existence of secondary peaks of emission after the maximum radio relic luminosity output. These secondary peaks are generated by subsequent galaxy cluster mergers producing new, often less luminous, radio shocks in the process. In general, cluster mergers are usually far from being idealised binary mergers as, usually, a plethora of merging substructures complicates the picture. Moreover, although most major mergers in massive clusters show fairly radial orbits, larger impact parameters are also likely to affect shock formation and their radio emission.  

We note that these light curves are modelled assuming that the lifetime of the CRes, which originate the observed radio emission, is short compared to the timescale of the cluster evolution. For longer electron lifetimes, as revealed by observations carried out at very low frequencies, the synchrotron emission of electrons will remain at significantly later times than the time of acceleration. This would eventually cause a longer duration of the non-thermal radio emission output. In this work, however, we do not model such effects.

\begin{figure}
    \hspace{-0.4cm}
    \includegraphics[width=0.99\columnwidth]{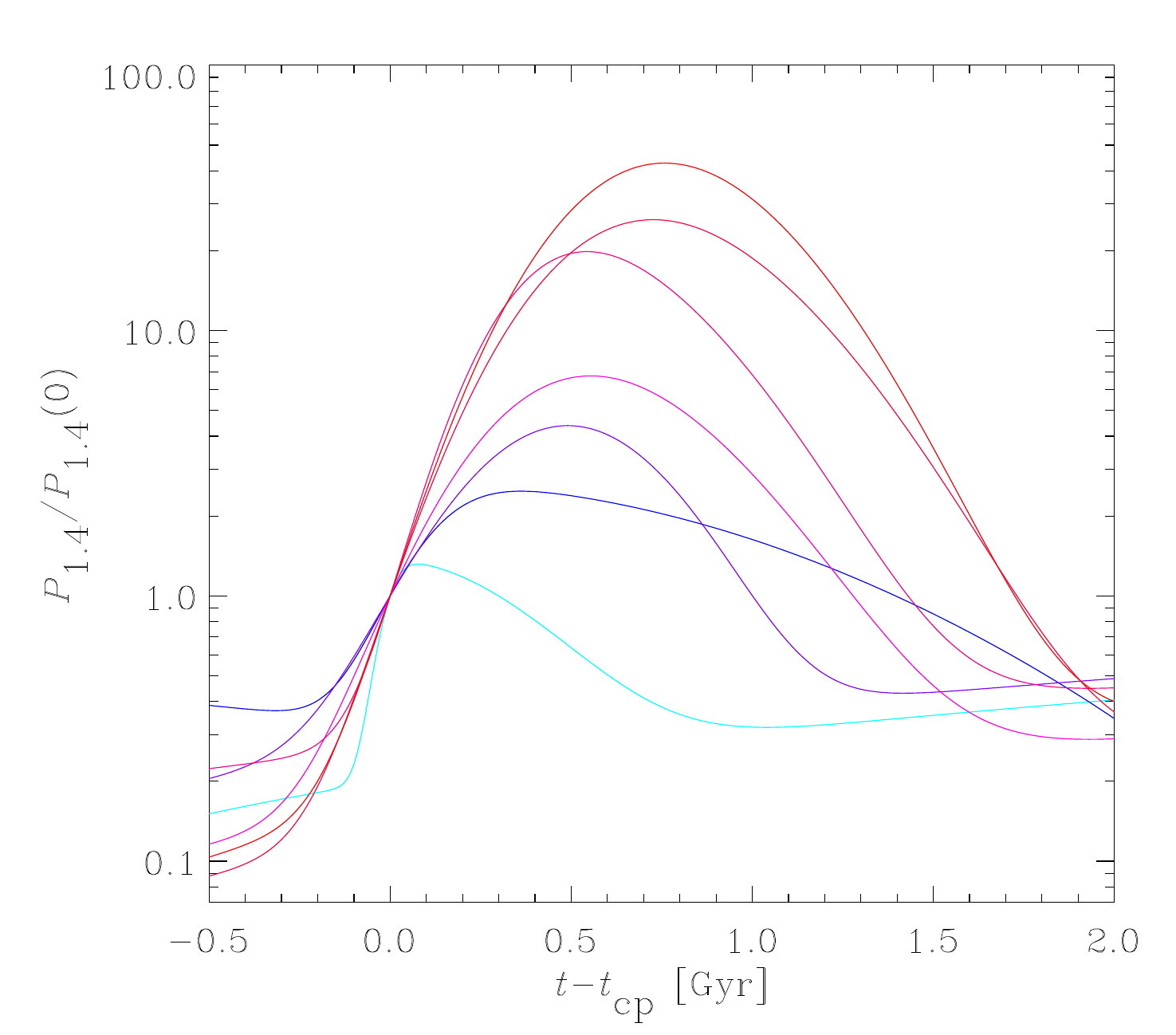}
    \caption{Best-fit median light curves at $1.4\,$GHz inside the $R_{200,1}(t)$ spheres normalised to the radio luminosity output at core-passage, $P_{1.4}(0)$, as a function of time relative to the core-passage time for galaxy group and cluster mergers. Different curves correspond to $M_{200,1}$ mass bins of 
    $[1,2.5),[2.5,5),[5,7.5),[7.5,10),[10,25),[25,50),[50,\infty)$ in units of $10^{13}\,{\rm M}_{\odot}$ (from light blue to red, respectively).}
    \label{fig:mean_normalisedLCs}
\end{figure}

\subsection{Average gischt radio power evolution}
\label{sec:average_lc}

In this section, we would like to study the `average' radio power of radio relics in our sample in order to quantify the rise and fall of radio luminosity during relic formation, considering median duration, peak luminosity and skewness, as a function of cluster mass and redshift. To parametrise the mass of a given merger we use, as before, $M_{200,1}$; that is, the mass of the main colliding cluster at $t_{\rm start}$, regardless of the $M_{200,2}$ value. As already mentioned, for every major merger in our catalogue, we estimate the core-passage time of the two most massive haloes and measure the time evolution of mergers with respect to the core-passage.

Fig.~\ref{fig:mean_radioLCs} shows the median radio power evolution at $1.4\,$GHz within $R_{200,1}(t)$ during galaxy cluster mergers in our synthetic sample relative to the core-passage time for all galaxy groups (solid red circles) and clusters (solid blue circles) in the catalogue. Since our sample mainly comprises major mergers \citep{Contreras22}, these statistics correspond to merger ratios of $\mu\gtrsim0.3$. To better characterise the data, a decile above and below the median values are shown as thin solid lines indicating that typical dispersion in radio outputs is large owing to the huge variability within the galaxy cluster zoo. To account for the asymmetry of the radio luminosity evolution, best-fitting curves adopting a skewed Gaussian function plus a linear baseline to mimic the cumulative effect of radio emission within the cluster volume are shown as solid lines. Specifically, for each light curve, we adopt

\begin{eqnarray}
        P_{1.4,{\rm fit}}(t)
        & \equiv &
        \frac{a_{0}}{\sqrt{2\pi}\sigma}\exp\left(-\frac{(t-t_{\rm cp}-t_0)^2}{2 \sigma^2}\right)
        \nonumber
        \\ &&
        \quad
        \times\left(1+\erf{\left(\gamma\frac{t-t_{\rm cp}-t_0}{\sqrt{2}\sigma}\right)}\right)
        \label{Eq:bestfit}
	\\ && 
        \quad
        +s_0 \left(t-t_{\rm cp}\right)+s_1{\rm ,}
        \nonumber
\end{eqnarray}

\noindent where the quantity $t-t_{\rm cp}$ is the cosmic time with respect to the core-passage. The background emission, given by the coefficients $s_0$ and $s_1$, may also include spurious radiation originating from other merging substructures and/or any other non-related shock fronts within the search volume. For completeness, the corresponding best-fit median light curves for all radiation measured within a radius of $2R_{200,1}(t)$ from the centre of the main progenitors are also shown in Fig.~\ref{fig:mean_radioLCs}. As already discussed, it can be seen that the peak of the light curve is already well captured within $R_{200,1}(t)$, with the main difference between the solid and dashed lines being the inclusion of more radio emission at later times in the latter. Therefore, in the following, we mainly focus on light curves within $R_{200,1}(t)$, unless otherwise stated. Best-fit parameters for the solid and dashed curves shown in Fig.~\ref{fig:mean_radioLCs} are presented in Table~\ref{tab:bestfit_param}.\footnote{{If an electron acceleration efficiency, $\xi'_{\rm e}$, different from the one introduced in Section~\ref{sec:radio_emission} were adopted, parameters $a_0, s_0$, and $s_1$ could easily be re-scaled after multiplying their best-fit values by $\xi'_{\rm e}/\xi_{\rm e}$ with $\xi_{\rm e}=5\times10^{-5}$.}}

To become independent of the, ultimately, arbitrary efficiency parameter, Fig.~\ref{fig:mean_normalisedLCs} shows the resulting best-fit median output evolution during mergers normalised to the radio luminosity at core-passage for $M_{200,1}$ average mass intervals of $1.75,3.75,6.25,8.75,17.5,37.5,\geq\,$$50$ in $10^{13}\,$M$_{\odot}$ units. With this choice, all mass bins have no less than $40$ cluster mergers in each bin, with the only exception of the more massive one which has 22 clusters with primary cluster masses up to about $10^{15}\,$M$_{\odot}$ and a mean mass of $6.95\times10^{14}\,$M$_{\odot}$.
This plot clearly shows that the merger-induced radio luminosity increase significantly after core-passage, from a factor up to $\sim10$ for group-like systems to about $\gtrsim10-50$ times for the most massive ones. However, individual light curves can show even higher-luminosity bursts, as is seen for example in some of the mergers shown in the right-hand panels of Fig.~\ref{fig:radioLCs}. Apart from the less massive bins, which show a moderate but extended emission after peak luminosity, the remaining mass bins in Fig.~\ref{fig:mean_normalisedLCs} show a similar shape of the luminosity evolution that tends to shift towards the right as mass increases. The similarity between the different curves highlights the fact that the same underlying physical phenomenon is taking place.    

As is seen in these plots, more massive mergers result in both larger luminosities and longer duration times of the radio-loud phase. According to Eq.~(\ref{eq:radio_power}), higher radio outputs can be explained mostly by the larger kinetic energies achieved by the secondary clusters during more massive cluster collisions. Our radio model implies that higher gas temperatures translate into more pronounced high-energy tails of the particle velocity distribution, providing more target electrons in shocks that can be accelerated by DSA \citep[][]{HB07}. This, in turn, will produce larger outbursts of non-thermal radio emission in the presence of magnetic fields in the ICM.  

Interestingly, for the most massive systems, some excess radio emission produced by small shock fronts can be seen even before core-passage, as already suggested from the snapshots of Fig.~\ref{fig:merger_evolution_example}. On the other hand, the radio emission bumps seen after the luminosity peak in the galaxy group and cluster samples -- at times $t-t_{\rm cp}\gtrsim1.6\,$Gyr (see Fig.~\ref{fig:mean_radioLCs}) -- result from merger shocks produced by subsequent galaxy cluster collisions, unrelated to the original ones. This feature is not uncommon during galaxy cluster mergers since realistic situations usually depart from the ideal, isolated binary merger scenario. In fact, we found that about $25\%$ of the mergers in our galaxy group and cluster catalogues are followed by other mergers produced by relatively massive substructures capable of boosting the radio luminosity after $t_{\rm peak}$, the time of peak radio luminosity. From visual inspection of the merger evolution of clusters in the sample, it can be seen that, in most cases, these radio bumps are produced by new, less-luminous radio shocks. As a consequence, these mergers produce a secondary peak of radio emission showing up $\gtrsim1-2\,$Gyr after the core-passage, demonstrating once again that the build up of galaxy clusters is not free from complexities, as is seen, for instance, in Fig.~\ref{fig:radioLCs}. An improved characterisation of radio shock formation, amplification and evolution in subsequent mergers will be studied in a future work.

\begin{figure}
    \hspace{-0.4cm}
    \includegraphics[width=0.99\columnwidth]{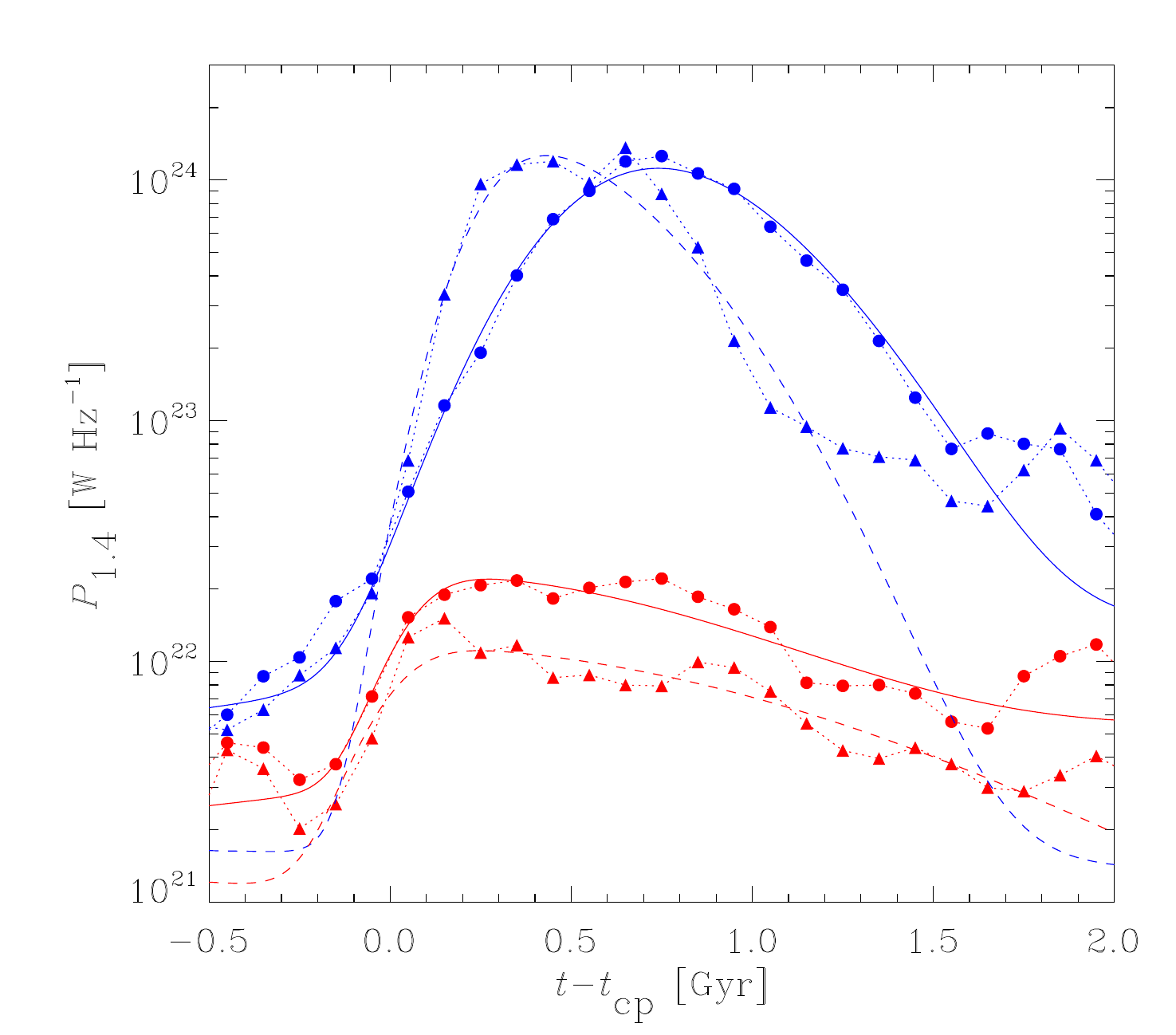}
    \caption{Median light curves at $1.4\,$GHz during merger events for $z<1$ (solid blue circles) and $z>1$ (solid red circles) as a function of time with respect to the core-passage. The median main progenitor cluster mass is $M_{200,1}=2.77\times10^{14}\,$M$_{\odot}$ and $2.34\times10^{13}\,$M$_{\odot}$, respectively. Solid triangles correspond, in each case, to mergers with impact parameters $b<200\,$kpc measured at $z_{\rm start}$. Fitting functions are shown as solid and dashed lines (see Section~\ref{sec:impact_param} for details).}
    \label{fig:mean_radioLCs2}
\end{figure}

\subsubsection{Early and late merger-induced radio light curves}

In Fig.~\ref{fig:mean_radioLCs2}, we split the merger sample into `early' ($z>z_{\rm cut}$) and `late' ($z<z_{\rm cut}$) mergers (solid red and blue circles, respectively) using $z_{\rm cut}=1$ because observed relics in current surveys are detected below this value. The redshift of each merger in our catalogue is computed as the mean value of $z_{\rm start}$ and $z_{\rm end}$. In this way, every merger redshift is considered after the merger process has already started and before the core-passage. We note, however, that this choice is somewhat arbitrary but it does not change the validity of our conclusions. As has already been seen in the right-hand panel of Fig.~\ref{fig:merger_rate}, most massive mergers take place at lower redshifts, whereas mergers dominated by galaxy groups are located at higher redshifts (see also Fig.~\ref{fig:M200_vs_z}). In fact, the median $M_{200,1}$ values for the early and late merger samples selected in this way give $2.34\times10^{13}$ and $2.77\times10^{14}\,$M$_{\odot}$, respectively, in line with our previous results in Section~\ref{sec:merger_rate}. The corresponding median $M_{200,2}$ value is $1.16\times10^{13}$ in the early sample and $1.72\times10^{14}\,$M$_{\odot}$ in the late sample. As a natural consequence of this redshift segregation, the early merger sample produces less luminous relics in comparison to the most massive, late mergers. For the acceleration efficiency adopted in Section~\ref{sec:radio_emission}, early (late) mergers reach typical peak luminosities of about $\log (P_{1.4}\,$W$^{-1}\,$Hz$)\sim 22\,(24)$. It is worth reminding that these luminosity estimates result from the calibration of simulated galaxy cluster samples to reproduce the observed number of clusters with relics in the NRAO VLA Sky Survey \citep[][]{Nuza17}, and therefore the resulting radio outputs in our model should be of the same order of magnitude as the observations. Nevertheless, it is out of the scope of this paper to provide a more detailed analysis of the relic radio power in mock galaxy cluster surveys. As in the previous section, best-fit parameters for the curves shown in Fig.~\ref{fig:mean_radioLCs2} (see solid lines) are presented in Table~\ref{tab:bestfit_param}. 

\subsubsection{Dependence with galaxy cluster impact parameter}
\label{sec:impact_param}

To study the effect of orbital parameters in the radio luminosity output of relics, we also computed the `impact parameter' of the two most massive merging cluster haloes using the relative position and velocity vectors at $t_{\rm start}$, the starting time of the merging process. The angle, $\theta_{21}$, formed by the axis joining the two colliding haloes and their relative velocity is therefore

\begin{equation}
    \cos(\theta_{21})=\frac{\mathbf{r}_{\rm rel}\cdot\mathbf{v}_{\rm rel}}{|\mathbf{r}_{\rm rel}||\mathbf{v}_{\rm vel}|},
\end{equation}

\noindent where $\mathbf{r}_{\rm rel}\equiv\mathbf{r}_2-\mathbf{r}_1$ and $\mathbf{v}_{\rm rel}\equiv\mathbf{v}_2-\mathbf{v}_1$ are the relative position and velocity of the secondary substructure with respect to the central halo, respectively. With this definition, two radially approaching substructures fulfill the condition $\cos(\theta_{21})=-1$, whereas systems in a circular orbit give $\cos(\theta_{21})=0$. The impact parameter, $b$, can then be computed as

\begin{equation}
    b=|\mathbf{r}_{\rm rel}|\tan(\pi-\theta_{21}).
\end{equation}

This formula gives a measure of the distance between the two merging haloes at core-passage by extrapolating the relative velocity vector from $t_{\rm start}$ to $t_{\rm cp}$. Similarly to the angle, $\theta_{21}$, the impact parameter, $b$, can then be used to characterise the orbit of the two main merging haloes with ideal impact parameters of $b=0$ and $b=\infty$ corresponding to purely radial and tangential orbits, respectively.

The median $\theta_{21}$ angle of the sample for all groups and clusters with $M_{\rm 200,1}\geq 10^{13}\,$M$_{\odot}$ is $\theta_{21}\approx160^{\circ}$ (i.e. the relative velocity and position vectors align within a cone of $\sim 20^{\circ}$). The resulting median impact parameter of the whole sample in this case is $b=255\,$kpc, which represents a median fraction of $b_{200}=0.56$ in $R_{200,1}$ units. We note, however, that the impact parameter must be defined using the relative position and velocity vectors at a specific time that, in the present work, we take as $t_{\rm start}$ (see Section~\ref{sec:merger_sample}). At this time, the median physical distance between subclusters in the complete galaxy group and cluster merger sample is $755\,$kpc. Similarly, if we consider the cluster merger sample only with $M_{200,1}\geq10^{14}\,$M$_{\odot}$ at $t_{\rm start}$, we obtain a median impact parameter of $485\,$kpc (i.e. $b_{200}=0.52$) and a median physical subcluster distance of $1436\,$kpc.

These results imply that most of the mergers happen in fairly radial orbits, although the tangential component is not completely insignificant. In general, our galaxy group and cluster sample shows that tangential orbits during mergers are rare, which can be understood by the fact that secondary cluster substructures merge onto the centrals from preferred directions determined by the large-scale filaments connected to the nodes of the cosmic density field \citep[see e.g.][]{Rost24}, although at cluster scales substructure orbits usually get randomised.

\begin{figure}
    \hspace{-0.3cm}
    \includegraphics[width=0.99\columnwidth]{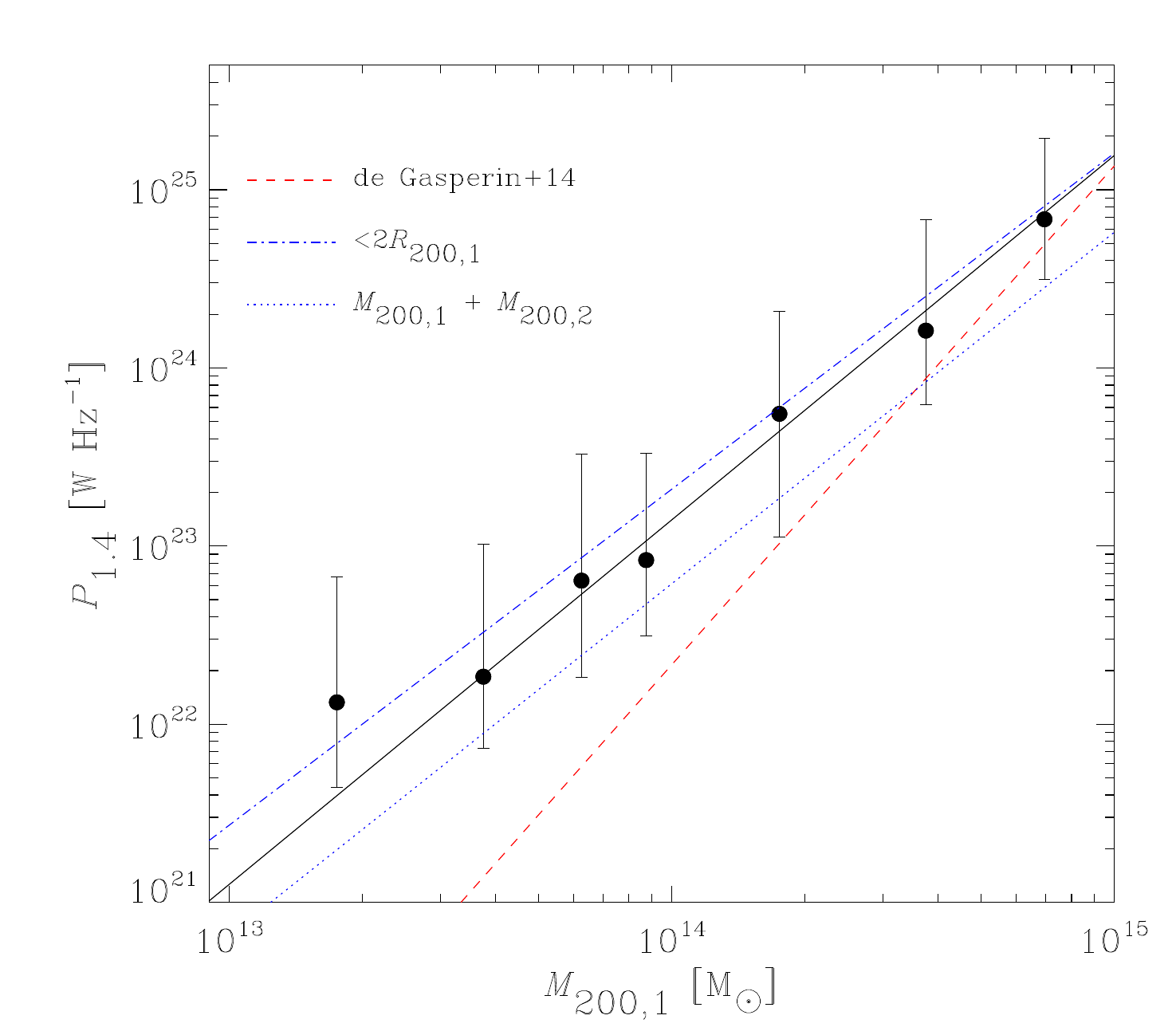}
    \caption{Peak radio power at $1.4\,$GHz versus mass of the main cluster progenitor during the simulated merger events (solid circles). The solid line corresponds to the best-fit linear relation $P_{1.4}=1.4\times10^{23}\,(M_{200,1}/10^{14}\,\rm{M}_{\odot})^{2.05}\,$W\,Hz$^{-1}$. Error bars indicate 20 and 80 percentiles in the radio luminosity outputs. Dash-dotted and dotted lines correspond to the two additional cases discussed in the text. For comparison, we also include the corresponding \cite{deGasperin14} observed relation corrected to $M_{200}$ (dashed line).}
    \label{fig:radiopower_mass}
\end{figure}

Fig.~\ref{fig:mean_radioLCs2} also shows the median light curves of cluster mergers with impact parameters $b<200\,$kpc for the early and late merger samples (solid red and blue triangles) together with the corresponding best-fitting curves (dashed lines). We note that, in each case, all fits were performed avoiding the contamination of subsequent mergers that naturally increase the output radio power after $\sim1-2\,$Gyr in some of the merging clusters (see Section~\ref{sec:average_lc}). For instance, this is evident in the $b<200\,$kpc, late merger sample at $t-t_{\rm cp}\gtrsim1.2\,$Gyr, where light curve data (solid blue triangles) departs from the best-fitting curve (blue dashed line) as a result of a few subsequent mergers in the sample. In general, best-fitting curves represent the radio luminosity evolution of the major group  and galaxy cluster mergers studied in this work, avoiding any secondary emission peak owing to smaller accreting substructures.

A moderate dependence of the radio luminosity evolution with the impact parameter is observed, mainly affecting the duration of the radio-loud phase of relics after peak luminosity. This trend is more pronounced in the case of late (i.e. more massive) mergers with light curves corresponding to less-radial mergers decaying more slowly. In particular, in the late merger sample, the median radio luminosity of more radial mergers (solid blue triangles) peaks a few hundred million years earlier than the full late sample. Similarly, in the former, the radio power decays faster, reaching about the same radio luminosity $\sim500\,$Myr earlier. In the case of early mergers, the trend is less pronounced but consistent with the results of the more massive clusters. These trends are likely related to the fact that less radial (i.e. more tangential) mergers tend to take longer timescales to reach a new state of dynamical equilibrium thus delaying radio emissions, whereas systems with more radial orbits immediately start relaxing after core-passage. A more thorough examination of these effects, however, needs to be performed and will be studied elsewhere. 

Finally, if we split the sample into mass bins we found that most massive mergers tend to have larger impact parameters. For galaxy groups and clusters in the average primary mass bins introduced in Section~\ref{sec:average_lc}, the resulting impact parameters at core-passage are $b=[113,149,257,258,351,405,444]\,$kpc in physical units. This increasing trend is naturally explained by the larger size of more massive galaxy cluster haloes. However, the `normalised' impact parameters in each cluster mass bin give values of $b_{200}=[0.63,0.51,0.68,0.61,0.50,0.44,0.57]$, which, conversely, shows an almost constant trend.

\subsection{Relic radio power--galaxy cluster mass correlation}

To study the dependence of the gischt radio luminosity peak with cluster mass we split the merger sample in $M_{200,1}$ mass intervals and plot the resulting correlation in Fig.~\ref{fig:radiopower_mass}. In this plot, we use the same primary cluster mass bins defined before, which let us study the radio power--mass relation including the less-massive mergers. The error bars indicate the variation in radio luminosities, measured by the 20 and 80 percentiles in the corresponding mass-dependent luminosity distributions. For a given cluster mass bin, typical dispersions span about one order of magnitude from the minimum to the maximum radio power values. The solid line corresponds to the best-fit power-law relation, $P_{1.4}\propto M_{200,1}^{2.05}$, to the simulated data (see caption of Fig.~\ref{fig:radiopower_mass}). Moreover, we also show the best-fit linear relations obtained computing all radiation (i) within $2R_{200,1}(t)$ (dash-dotted line) and (ii) within $R_{200,1}(t)$ but using $M_{200,1}+M_{200,2}$ for the mass binning (dotted line). This last case is considered to assess the effect of the secondary cluster in the radio--power mass relation. The best fits are $P_{1.4}=2.09\times10^{23}\,(M_{200,1}/10^{14}\,{\rm M}_{\odot})^{1.88}\,$W\,Hz$^{-1}$ in case (i) and $P_{1.4}=6.15\times10^{22}\,(M_{200,+}/10^{14}\,{\rm M}_{\odot})^{1.97}\,$W\,Hz$^{-1}$ in case (ii) where $M_{200,+} \equiv M_{200,1}+M_{200,2}$. For reference, we also include the observed trend by \cite{deGasperin14} after correcting cluster masses to $M_{200}$ assuming an NFW profile and halo concentrations from \cite{Duffy08}.

In line with the discussion of Section~\ref{sec:average_lc}, there is an evident increase in radio luminosity with cluster mass that determines a correlation spanning almost 2 orders of magnitude, from galaxy groups of about $10^{13}\,$M$_{\odot}$ to the most massive clusters. This result is consistent with previous findings using a set of simulated massive clusters from the MUSIC-2 sample \citep[][]{Nuza17} and, recently, from the TNG-Cluster simulation \citep{Lee24}. Nevertheless, in these works, the low-mass end of the merging cluster masses have higher values of about $\log(M_{200,1}/\rm{M}_{\odot})\sim 14.2$ \citep{Lee24} and $\sim14.7$ \citep{Nuza17}, whereas here we expand the radio power-mass correlation all the way down to galaxy group scales.
For the TNG-Cluster simulation, \cite{Lee24} found that the slope of the average power-law relation is about $1.5$. However, after rescaling it by weighting clusters with the number ratio of the simulated and theoretical mass functions to compensate for their lack of low-mass clusters, they obtained a corrected slope of about $2$, which is in excellent agreement with our results. In particular, this slope does not significantly change if we consider, for example, a larger search radius as in case (i), or an alternative mass binning as in case (ii). This is true because we are mainly focusing on the luminosity peaks of the median light curves which, in comparison to the fiducial correlation shown by the solid line, are similar or simply shift the curve to the right, respectively. The resulting slopes are not far off from the theoretical expectation ($P_{\nu}\propto M_{\Delta}^{5/3}$, where $\nu$ is the frequency and $\Delta$ is the mass density contrast) readily derived from scaling arguments \citep[see e.g.][]{deGasperin14}.

\begin{figure}
    \hspace{-0.4cm}
    \includegraphics[width=0.99\columnwidth]{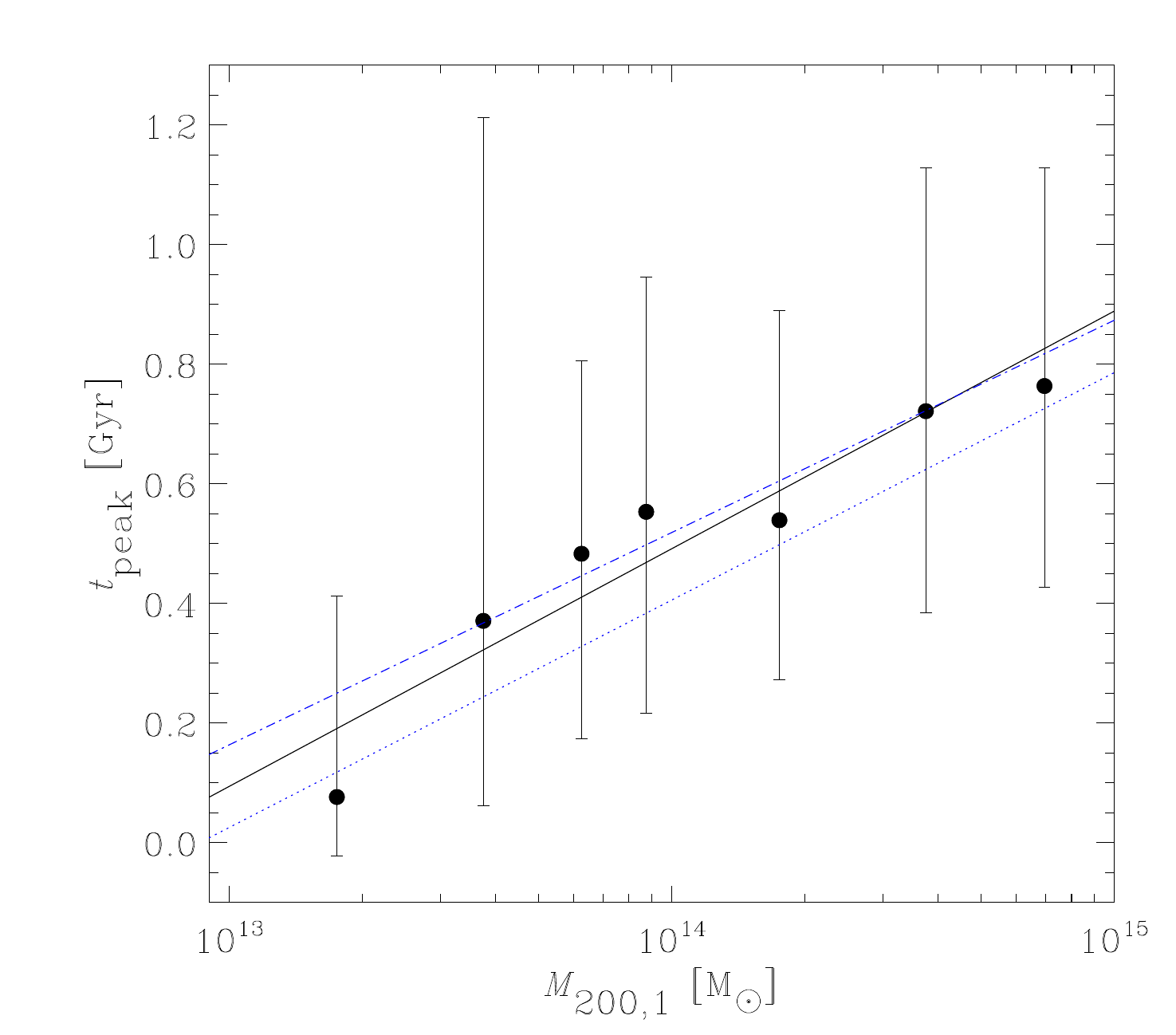}
    \caption{Luminosity peak time of the corresponding radio light curves in each cluster mass bin versus mass of the main cluster progenitor. The solid line corresponds to the best-fit $t_{\rm peak}=(0.39\times\log(M_{200,1}/10^{14}\,\rm{M}_{\odot})+0.49)~$Gyr. Bars show the FWHM emission time, providing a measure of relic duration for each cluster mass bin. Dash-dotted and dotted lines correspond to the additional best-fitting cases shown in Fig.~\ref{fig:radiopower_mass}.}
    \label{fig:radioduration_mass}
\end{figure}

Interestingly, our simulated correlation ($P_{1.4}\propto M_{200,1}^{2.05}$) is somewhat flatter than the \cite{deGasperin14} observed relation ($P_{1.4}\propto M_{200,1}^{2.8}$). This could be related to the fact that distant relics are preferentially observed if their intrinsic luminosity is higher, boosting the slope of the relation at large cluster masses, as has been shown by \cite{Nuza17} using mock observations of the MUSIC-2 galaxy cluster sample. We also stress the fact that this correlation is determined by the peak relic luminosity, whereas, as has been demonstrated by \cite{Nuza17} \citep[but see also][]{Lee24}, there is a large spread of luminosities for a given cluster mass, indicating that the specific details of each merging process such as orbital parameters (see previous section) and local thermodynamical properties of the gas also play a role in establishing the gischt luminosity outputs.

\subsection{Relic duration versus galaxy cluster mass}

Another interesting aspect that deserves attention is the duration of the radio-loud phase of merger-induced shock fronts causing relics. As is seen in Figs.~\ref{fig:mean_normalisedLCs} and~\ref{fig:mean_radioLCs2}, radio luminosities rise abruptly after core-passage increasing up to about 2 orders of magnitude at $t_{\rm peak}$ for the most massive cluster mergers. Overall, the light curves span a total time of about $1$--$2\,$Gyr from relic onset at $t-t_{\rm cp}\sim0$ to demise at $t-t_{\rm cp}\gtrsim2\,t_{\rm peak}$. The peak luminosity time of mergers after core-passage for groups and clusters in our sample is $t_{\rm peak}\sim0.1$--$0.8\,$Gyr. The latter can be compared with typical shock crossing times in galaxy clusters defined by $t_{\rm cross}\sim l/v_{\rm sh}$, where $l$ is the size of a galaxy cluster and $v_{\rm sh}$ is the shock velocity. Assuming typical values of $l\sim1000\,$kpc and $v_{\rm sh}\sim1000\,$km\,s$^{-1}$ this results in an estimate for the shock crossing time of $t_{\rm cross}\sim1\,$Gyr, which goes in line with the results from simulations.   

Fig.~\ref{fig:radioduration_mass} shows a measure of relic timescales characterised by the peak luminosity time as a function of $M_{200,1}$ assuming the same set of primary cluster mass bins than in the previous figure. Bars are computed from times where the median radio power is higher than half of the maximum best-fit model luminosity and provide a measure of relic duration from onset to demise. The upper bars are sometimes longer than the lower ones, which highlights the asymmetric nature of the gischt radio light curves. Overall, a moderate dependence of $t_{\rm peak}$ with $M_{200,1}$ is observed (highlighted by the best-fit relation), with more massive mergers lasting about $2$--$3$ times longer than less massive ones, essentially owing to the larger sizes of more massive clusters. When considering the additional cases (i) and (ii) discussed in the previous section we obtain similar relations that are contained well within relic duration times and therefore do not affect our conclusions. If we disregard galaxy group mergers and focus only on typical cluster scales, with main progenitor masses $M_{200,1}\gtrsim 10^{14}\,$M$_{\odot}$, the correlation with mass becomes weaker, and $t_{\rm peak}$ is in the range of approximately $0.6$ to $0.8\,$Gyr.

\section{Summary}
\label{sec:summary}

In this work, we have studied the evolution of gischt radio relic luminosity in a sample of 279 simulated galaxy group and cluster mergers with masses (defined at the merger starting time, $t_{\rm start}$) above $M_{200,1}\gtrsim10^{13}\,$M$_{\odot}$ and a median main cluster mass of $M_{200,1}=8.76\times10^{13}\,$M$_{\odot}$ obtained from \textsc{The Three Hundred} project, a suite of cosmological simulations of high-density regions in the cosmic web. Shocked gas regions produced during galaxy cluster collisions were identified using the technique adopted in \cite{Nuza17} to model the non-thermal radio emission output of accelerated electron populations according to the radio emission model of \cite{HB07}. Magnetic field magnitudes were modelled by following \cite{Nuza12,Nuza17} and using a power-law relation with local electron density in agreement with scalings obtained both from observations and cosmological simulations. For every simulated volume in our sample, we built the radio luminosity light curve of the most massive galaxy cluster in each region during its complete cosmic history. From these light curves, galaxy cluster merger times were identified using the merger catalogue of \cite{Contreras22} in order to determine the core-passage time of the two main progenitors to characterise radio luminosity outputs during merger events and study `average' properties as a function of the cluster mass, merger ratio, impact parameter, and merger redshift. In particular, since our cluster merger catalogue mainly comprises major mergers, we have restricted our analysis to merger mass ratios of at least $\mu\gtrsim 0.3$.
\\

\noindent
The main findings of this work are summarised as follows: 

\begin{itemize}
    \item Our merger catalogue indicates that less massive mergers involving galaxy groups with $M_{200,i}\sim \rm{few}\times10^{13}\,$M$_{\odot}$ are generally taking place at redshifts of $z\sim2$, whereas more massive cluster mergers with $M_{200,i}\sim \rm{few}\times10^{14}\,$M$_{\odot}$ tend to peak at smaller redshifts of $z\lesssim 0.5$, in agreement with the hierarchical nature of halo assembly. However, some overlap between our galaxy group and cluster merger counts exists in the redshift range of $z\sim0.5-1.5$. 
      
    \item The redshift peak of the most massive cluster merger sample is consistent with the mean observed redshift distribution of relics; that is, $z\sim0.3$ \citep[][]{Jones23}. Therefore, our cosmological cluster merger scenario suggests that observed radio relics are generally produced by the most massive merger events in the Universe.  

    \item We do not find any major merger without an increased radio emission in the aftermath. Every major merger causes a radio relic. For the electron efficiency parameter, $\xi_{\rm{e}}$, assumed in this work (see Section~\ref{sec:radio_emission}), typical gischt luminosities at $1.4\,$GHz for the galaxy group ($10^{13}<M_{200,1}/{\rm M}_{\odot}<10^{14}$) and cluster ($M_{200,1}\geq10^{14}\,{\rm M}_{\odot}$) samples are $P_{1.4}\sim10^{22}$ and $\sim10^{24}\,$W\,Hz$^{-1}$, respectively. These values, however, can be scaled accordingly if a different efficiency parameter was adopted.

    \item Most of the group and cluster mergers in the sample have fairly radial orbits with relative position and velocity vectors of the two most massive colliding structures aligning at $t_{\rm start}$ within a cone with a median opening angle of $\sim20^{\circ}$ and a median impact parameter of $b=255\,$kpc in physical coordinates ($b_{200}=0.56$ in $R_{200,1}$ units). If we only consider galaxy clusters with $M_{200,1} \geq 10^{14}\,$M$_{\odot}$, we obtain $b=485\,$kpc or $b_{200}=0.52$ in normalised units.
     
    \item We find significant variations in the radio luminosity of each cluster studied. The radio light curves over cosmic time are characterised by one or more peaks. Interestingly, around $25\%$ of the systems show a secondary peak generally found in the aftermath of large merger events owing to other, usually minor, subsequent mergers. This is in excellent agreement with the commonly accepted scenario that merger events generate shock waves that lead to observable radio emission. 
      
    \item When splitting the merger sample in differential mass bins of the primary clusters, we found that merger-induced non-thermal emission from shocks increase significantly after core-passage, from a factor of a few for group-like systems in the range $M_{200,1} \sim 1$--$10\times10^{13}\,$M$_{\odot}$ to about $10$--$50$ times for the most massive clusters above $\sim 10^{14}\,$M$_{\odot}$.
      
    \item The median radio luminosity within $R_{200,1}(t)$ at $t_{\rm peak}$ scales with the mass of the main progenitor, $M_{200,1}$. The resulting (fiducial) best-fit correlation after considering all mergers is $P_{1.4,\rm{peak}}=1.4\times10^{23}\,(M_{200,1}/10^{14}\,\rm{M}_{\odot})^{2.05}\,$W\,Hz$^{-1}$ for a global radio power normalisation governed by $\xi_{\rm e}$, spanning almost two orders of magnitude in mass from galaxy group scales to the most massive clusters. If, instead, we plot the radio power as a function of $M_{200,1}+M_{200,2}$, the slope remains essentially unchanged. If, alternatively, we measure all radiation within $2R_{200.1}(t)$ as a function of $M_{200,1}$, we obtain a similar correlation but with a somewhat higher normalisation than the fiducial one and a slope of $1.88$.
      
    \item We have found a trend of $t_{\rm peak}$ with the mass of the primary merging system, $M_{200,1}$. In general, radio luminosity peaks at $\sim 0.1-0.8\,$Gyr after core-passage for all mergers in the sample. The best-fit relation in the fiducial case is $t_{\rm peak}=(0.39\times\log(M_{200,1}/10^{14}\,\rm{M}_{\odot})+0.49)~$Gyr. In the latter, the full width at half maximum (FWHM) emission time does not change noticeably with cluster mass, except for the $[2.5,5)\times10^{13}\,$M$_{\odot}$ mass bin, where it spans a time interval from $\sim 100\,$Myr to $\sim1.2\,$Gyr, reflecting more disturbed merger scenarios at this mass range. Finally, if we mainly focus on the high-mass end ($M_{200,1}\gtrsim 10^{14}\,$M$_{\odot}$) -- typical of galaxy clusters -- the correlation gets milder with $t_{\rm peak}$ in the range $\sim0.6-0.8\,$Gyr.
\end{itemize}

The simulations show that cluster mergers are generally accompanied by shock waves that propagate through the ICM to the cluster periphery. Radio relic observations suggest that these shock fronts are often radio loud, indicating that part of the dissipated energy is channelled into a non-thermal component in the ICM. In turn, this can affect the pressure in the cluster periphery even after the shock front has propagated further out. Additional measurements, such as IC emission from the shock region to disentangle CRe density and magnetic fields, and further modelling, are needed to quantify the relative contribution of the non-thermal and turbulent pressure components.

\begin{acknowledgements}
We are grateful with the anonymous referee for a constructive report that helped to improve this paper. SEN is member of the Carrera del Investigador Científico of CONICET. He acknowledges support from CONICET (PIBAA R73734), Agencia Nacional de Promoci\'on Cient\'{\i}fica y Tecnol\'ogica (PICT 2021-GRF-TI-00290) and UBACyT (20020170100129BA).  
AC-S, AK and GY acknowledge partial financial support from MICIU/FEDER (Spain) through  research grant number PID2021-122603NB-C21. Additional financial support comes from the European Union’s Horizon 2020 Research and Innovation programme under the Marie  Sk\l{}odowskaw-Curie grant agreement number 734374, i.e. the LACEGAL project\footnote{\url{https://cordis.europa.eu/project/rcn/207630\_en.html}}. This
work has been made possible by the ``The Three Hundred''
collaboration\footnote{\url{https://www.the300-project.org}}. The simulations used in this paper have been performed in the MareNostrum Supercomputer at the Barcelona Supercomputing Center, thanks to CPU time granted by the Red Española de Supercomputación. For the purpose of open access, the author has applied a creative commons attribution (CC BY) to any author accepted manuscript version arising.
\end{acknowledgements}

\bibliographystyle{aa}
\bibliography{refs}

\end{document}